\documentclass[letterpaper,twocolumn,10pt]{article}
\usepackage{usenix,epsfig,endnotes}

\usepackage{graphicx}
\usepackage{url}
\usepackage{syntax}
\usepackage{newfloat}
\usepackage{multicol}
\usepackage{hyperref}

\newcommand{\Comment}[1]{}

\newcommand{\SmallSpace}{\vspace*{-1.5ex}}

\newcommand{\para}[1]{\smallskip\noindent{\bf {#1}:}}

\DeclareFloatingEnvironment[
  fileext   = logr,
  listname  = {List of Grammars},
  name      = Grammar,
  placement = !htbp
]{BNF}

\newcommand{\distance}{5pt}
\usepackage{acronym}
\usepackage[ruled,noend]{algorithm2e}
\usepackage{mathtools}
\usepackage{url}
\usepackage{multirow}
\usepackage{mdwlist}
\usepackage{xspace}
\usepackage{amsmath}
\usepackage{amssymb}
\usepackage{misc}
\usepackage{subcaption}
\usepackage{listings}
\usepackage{enumerate}

\usepackage{enumitem}
\usepackage{adjustbox}
\usepackage{mdframed}

\usepackage{xcolor}

\usepackage[capitalize,nameinlink]{cleveref}
\hypersetup{%
	bookmarksnumbered, bookmarksopen=true, bookmarksopenlevel=1,%
}

\definecolor{mygreen}{rgb}{0,0.6,0}
\definecolor{mygray}{rgb}{0.5,0.5,0.5}

\crefname{figure}{Figure}{Figures}
\crefname{listing}{Query}{Queries}
\crefname{section}{Section}{Sections}
\crefname{table}{Table}{Tables}
\crefname{BNF}{Grammar}{Grammars}
\crefname{algorithm}{Algorithm}{Algorithms}

\SetKwInput{KwInput}{Input}
\SetKwInput{KwOutput}{Output}
\SetKwProg{Fn}{Function}{}{end}

\SetCommentSty{mycommfont}

\newcommand{\incode}[1]{\lstinline{#1}}

\newcommand{\dsl}{\textsc{Saql}\xspace}
\newcommand{\eat}[1]{}

\newcommand{\eg}{{\it e.g.,}\xspace}

\newcommand{\ie}{{\it i.e.,}\xspace}

\usepackage[labelfont=bf,skip=0pt]{caption}
\newcommand{\dist}{2pt}
\acrodef{CEP}{Complex  Event  Processing}
\acrodef{SAQL}{\emph{Stream-based Anomaly Query Language}}

\acrodef{APT}{Advanced Persistent Threats}
\acrodef{SIEM}{Security Information and Event Management}
\acrodef{IDS}{Intrusion Detection Systems}
\acrodef{DSL}{Domain Specific Language}

\eat{
\fancyhf{} 
\fancyhead[C]{Anonymous Submision \#9999 to ACM CCS 2017} 
\fancyfoot[C]{\thepage}

\setcopyright{none} 
\acmConference[Anonymous Submission to ACM CCS 2017]{ACM Conference on Computer and Communications Security}{Due 19 May 2017}{Dallas, Texas}
\acmYear{2017}

\settopmatter{printacmref=false, printccs=true, printfolios=true} 

}

\begin{document}
\date{}

\title{\Large \bf \dsl: A Stream-based Query System for Real-Time Abnormal System Behavior Detection}

\author{
	{\rm Peng Gao$^1$}
	\and
	{\rm Xusheng Xiao$^2$}
		\and
	{\rm Ding Li$^3$}
	\and
	{\rm Zhichun Li$^3$}
	\and
	{\rm Kangkook Jee$^3$}
	\and
	{\rm Zhenyu Wu$^3$}
	\and
	{\rm Chung Hwan Kim$^3$}
	\and
	{\rm Sanjeev R. Kulkarni$^1$}
	\and
	{\rm Prateek Mittal$^1$}
	\and
	{\normalsize $^1$Princeton University\; $^2$Case Western Reserve University\; $^3$NEC Laboratories America, Inc.}
	\and
	{\small $^1$\{pgao,kulkarni,pmittal\}@princeton.edu\; $^2$xusheng.xiao@case.edu\; $^3$\{dingli,zhichun,kjee,adamwu,chungkim\}@nec-labs.com} 
}

\maketitle

\thispagestyle{empty}
\pagestyle{empty}


\begin{abstract}
Recently, advanced cyber attacks, which consist of a sequence of steps that involve many vulnerabilities and hosts, compromise the security of many well-protected businesses.
This has led to the solutions that ubiquitously monitor system activities in each host (big data) as a series of events, and search for anomalies (abnormal behaviors) for triaging risky events.
Since fighting against these attacks is a time-critical mission to prevent further damage, these solutions face challenges 
in incorporating \emph{expert knowledge} to perform \emph{timely anomaly detection} over the large-scale provenance data.

To address these challenges, we propose a novel stream-based query system that takes as input, a real-time event feed aggregated from multiple hosts in an enterprise, and provides an anomaly query engine that queries the event feed to identify abnormal behaviors based on the specified anomalies.
To facilitate the task of expressing anomalies based on expert knowledge,
our system provides a domain-specific query language, \dsl, 
which allows analysts to express models for (1) \emph{rule-based anomalies}, (2) \emph{time-series anomalies},
(3) \emph{invariant-based anomalies},
and (4) \emph{outlier-based anomalies}.
We deployed our system in 
NEC Labs America comprising 150 hosts and evaluated it using 1.1TB of real system monitoring data (containing 3.3 billion events).
Our evaluations on a broad set of attack behaviors and micro-benchmarks show that our system has a low detection latency ($<$2s) and a high system throughput (110,000 events/s; supporting $\sim$4000 hosts), and is more efficient in memory utilization than the existing stream-based complex event processing systems.

\end{abstract}


\section{Introduction}

Advanced cyber attacks and data breaches plague even the most protected companies~\cite{ebay, opm, homedepot, target, equifax}. 
The recent massive Equifax data breach~\cite{equifax} has exposed the sensitive personal information of 143 million US customers. 
Similar attacks, especially in the form of advanced persistent threats (APT), are being commonly observed.
These attacks consist of \emph{a sequence of steps} across \emph{many hosts} that exploit different types of vulnerabilities to compromise security~\cite{tc,aptfireeye,aptsymantec}.

To counter these attacks, approaches based on \emph{ubiquitous system monitoring} have emerged as an important solution for actively searching for possible anomalies, then to quickly triage the possible significant risky events~\cite{backtracking,backtracking2,taser,taserdb, intrusionrecovery,mpi,introsec,lee2013high}. 
System monitoring observes \emph{system calls} at the kernel level to collect information about system activities.
The collected data from system monitoring facilitates the detection of abnormal system behaviors~\cite{anomalysurvey,idsbook}.

\begin{figure*}
	\centering
	\includegraphics[width=0.9\textwidth]{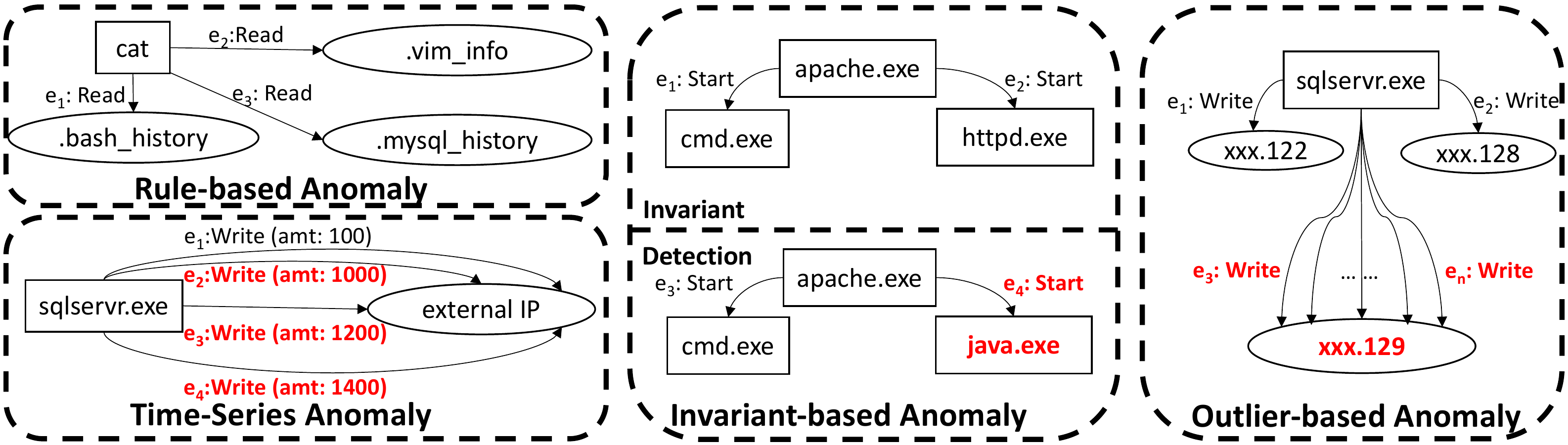}
	\caption{Major types of abnormal system behaviors ($e_1,\ldots,e_n$ are shown in ascending temporal order.)}
	\label{fig:moti}
\end{figure*}

However, these approaches face challenges in detecting multiple types of anomalies using system monitoring data. 
First, fighting against attacks such as APTs is a time-critical mission. 
As such, we need a \emph{real-time} anomaly detection tool to search for a ``needle in a haystack'' for preventing additional damage and for system recovery. 
Second, models derived from data have been increasingly used in detecting various types of risky events~\cite{idsbook}. 
For example, system administrators, security analysts and data scientists have extensive domain knowledge about the enterprise, including expected system behaviors.  
A key problem is how we can provide a real-time tool to detect anomalies while \emph{incorporating the knowledge} from system administrators, security analysts and data scientists?
Third, system monitoring produces huge amount of daily logs ($\sim$50GB for 100 hosts per day)~\cite{loggc,reduction}.
This requires \emph{efficient} real-time data analytics on the large-scale provenance data.

Unfortunately, none of the existing stream-based query systems and anomaly detection systems~\cite{streamanalytics4,cql,streamanomaly1,lee2013high} provide a comprehensive solution 
that addresses all these three challenges. 
These systems focus on specific anomalies and are optimized for general purpose data streams,
providing limited support for users to specify anomaly models by incorporating domain knowledge from experts.

\para{Contributions}
We design and build a novel stream-based real-time query system.
Our system takes as input a real-time event feed aggregated from multiple hosts in an enterprise, and provides an anomaly query engine.
The query engine provides a novel interface for users to submit anomaly queries using our \emph{domain-specific language},
and checks the events against the queries to detect anomalies in real-time.

\para{Language}
To facilitate the task of expressing anomalies based on domain knowledge of experts, our system provides a domain-specific query language, \emph{Stream-based Anomaly Query Language (\dsl)}.
\dsl provides (1) the syntax of event patterns to ease the task of specifying relevant system activities and their relationships,
which facilitates the specification of \emph{rule-based anomalies};
(2) the constructs for \emph{sliding windows} and \emph{stateful computation}
that allow stateful anomaly models to be computed in each sliding window over the data stream,
which facilitates the specification of \emph{time-series anomalies}, \emph{invariant-based anomalies},
and \emph{outlier-based anomalies} (more details in \cref{subsec:saqlquery}).
The specified models in \dsl are checked using \emph{continuous queries} over unbounded streams of system monitoring data~\cite{cql},
which report the detected anomalies continuously.

Rule-based anomalies allow system experts to specify rules to detect known attack behaviors or enforce enterprise-wide security policies.
\cref{fig:moti} shows an example rule-based anomaly, where a process (\incode{cat}) accesses multiple command log files in a relatively short time period, indicating an external user trying to probe the useful commands issued by the legitimate users.
To express such behavior, \dsl uses event patterns to express each activity in the format of \emph{\{subject-operation-object\}}
(\eg \incode{proc p1 write file f1}),
where system entities are represented as subjects (\incode{proc p1}) and objects (\incode{file f1}),
and interactions are represented as operations initiated by subjects and targeted on objects.

Stateful computation in sliding windows over a data stream enables the specification of stateful behavior models for detecting abnormal system behaviors such as time-series anomalies,
which lack support from existing stream query systems that focus on general data streams~\cite{streamanalytics4,streamanomaly1,MillWheel,streamanalytics3}.
\cref{fig:moti} shows a time-series anomaly, where a process (\incode{sqlservr.exe}) transfers abnormally large amount of data starting from $e_2$.
To facilitate the detection of such anomalies,
\dsl provides constructs for \emph{sliding windows} that break the continuous data stream into fragments with common aggregation functions (\eg \incode{count}, \incode{sum}, \incode{avg}).
Additionally, \dsl provides constructs to define \emph{states in sliding windows} and allow accesses to the states of past windows.
These constructs facilitate the comparison with historical states and the computation of moving averages such as three-period simple moving average (SMA)~\cite{hamilton1994time}.

Built upon the states of sliding windows, \dsl provides high-level constructs to facilitate the specification of invariant-based and outlier-based anomalies.
Invariant-based anomalies capture the invariants during training periods as models, and use the models later to detect anomalies.
\cref{fig:moti} shows an invariant-based anomaly, where a process (\incode{apache.exe}) starts an abnormal process (\incode{java.exe}) that is unseen during the training period.
\dsl provides constructs to define and learn the invariants of system behaviors in each state computed from a window,
which allow users to combine both states of windows and invariants learned under normal operations to detect more types of abnormal system behaviors.

Outlier-based anomalies allow users to identify abnormal system behavior through peer comparison, \eg finding outlier processes by comparing the abnormal processes with other peer processes.
\cref{fig:moti} shows an outlier-based anomaly, where a process (\incode{sqlservr.exe}) transfers abnormally larger amount of data to an IP address than other IP addresses.
\dsl provides constructs to define which information of a state in a sliding window forms a point and compute clusters to identify outliers.
The flexibility and extensibility introduced by \dsl allows users to use various clustering algorithms for different deployed environments.

\para{Execution Engine}
We build the query engine on top of Siddhi~\cite{siddhi} to leverage its mature stream management engine.
Based on the input \dsl queries, our system synthesizes Siddhi queries to match data from the stream,
and performs stateful computation and anomaly model construction to detect anomalies over the stream.
One major challenge faced by this design is the scalability in handling multiple concurrent anomaly queries over the large-scale system monitoring data.
Typically, different queries may access different attributes of the data using different sliding windows.
To accommodate these needs, the scheme employed by the existing systems, such as Siddhi, Esper, and Flink~\cite{siddhi,esper,flink}, is to make copies of the stream data and feed the copies to each query, allowing each query to operate separately.
However, such scheme is not efficient in handling the big data collected from system monitoring.

To address this challenge, we devise a master-dependent-query scheme that identifies compatible queries and groups them to use a single copy of the stream data to minimize the data copies.
Our system first analyzes the submitted queries with respect to the \emph{temporal dimension} in terms of their sliding windows
and the \emph{spatial dimension} in terms of host machines and event attributes.
Based on the analysis results, our system puts the \emph{compatible queries} into groups,
where in each group, a \emph{master query} will directly access the stream data and the other \emph{dependent queries} will leverage the intermediate execution results of the master query.
Note that such optimization leverages both the characteristics of the spatio-temporal properties of system monitoring data and the semantics of \dsl queries,
which would not be possible for the queries in general stream-based query systems~\cite{siddhi,esper,cql,flink}.


\para{Deployment and Evaluation}
We built the whole \dsl system (around 50,000 lines of Java code) based on the existing system-level monitoring tools (i.e., auditd~\cite{auditd} and ETW~\cite{etw}) and the existing stream management system (i.e., Siddhi~\cite{siddhi}). 
We deployed the system in 
NEC Labs America comprising 150 hosts.
We performed a broad set of attack behaviors in the deployed environment, and evaluated the system using 1.1TB of real system monitoring data (containing 3.3 billion events):
(1) our case study on four major types of attack behaviors (17 \dsl queries) shows that our \dsl system has a low alert detection latency ($<$2s);
(2) our pressure test shows that our \dsl system has a high system throughput (110000 events/s) for a single representative rule-based query that monitors file accesses, and can scale to $\sim$4000 hosts on the deployed server;
(3) our performance evaluation using 64 micro-benchmark queries shows that our \dsl system is able to efficiently handle concurrent query execution and achieves more efficient memory utilization compared to 
Siddhi, achieving 30\% average saving. 
All the evaluation queries are available on our \emph{project website~\cite{saql}}.

\eat{
To demonstrate the effectiveness of the \dsl system in expressing anomalies based on domain knowledge, 
we conduct a case study on four major attack behaviors: Outlook APT attack, SQL injection attack, Bash shellshock command injection attack, and suspicious system behaviors such as command history probing.
We performed these attacks in the deployed environment and collected 1.1TB of real system monitoring data. We compose 17 \dsl queries to detect the behaviors in these attacks and measure the system performance 
in terms of throughput and latency.
Moreover, to demonstrate the effectiveness of the master-dependent-query scheme employed by our \dsl system, we compare the performance of our \dsl system and Siddhi on three scenarios where multiple queries are concurrently executed to detect anomalies.
In summary, our evaluation results show that: (1) the \dsl system is able to promptly and efficiently detect a wide category of attacks with low latency (mostly $< 1\; ms$ plus $\sim1.5\; minutes$ end-to-end event collection, modeling, and transmission), and is able to efficiently support an enterprise comprising 100 hosts with 2500-3000 events/second as event throughput; (2) the master-dependent-scheme employed by the \dsl system is able to significantly reduce the need of data copies. Compared to Siddhi, our system achieves around 4x CPU savings and 10x memory space savings.
}


\section{Background and Examples}
\label{sec:motivation}
In this section, we first present the background on system monitoring
and then show \dsl queries to demonstrate the major types of anomaly models supported by our system.
The point is not to assess the quality of these models, but to provide examples of language constructs that are essential in specifying anomaly models, which lack good support from existing query tools.

\subsection{System Monitoring}
System monitoring data represents various system activities in the form of events along with time~\cite{backtracking,backtracking2,taser,wormlog}.
Each event can naturally be described as a system entity (subject) performing some operation on another system entity (object).
For example, a process reads a file or a process accesses a network connection.
An APT attack needs multiple steps to succeed, such as target discovery and data exfiltration, as illustrated in the cyber kill
chain~\cite{killchain}. Therefore, multiple attack footprints might be left as ``dots'', which can be captured
precisely by system monitoring.

System monitoring data records system audit events about the system calls that are crucial in security analysis~\cite{backtracking,backtracking2,taser,wormlog}.
The monitored system calls are mapped to three major types of system events:
(1) process creation and destruction, (2) file access, and (3) network access.
Existing work has shown that on main-stream operating systems (Windows, Linux and OS X), system entities in most cases are files, network connections and processes~\cite{backtracking,backtracking2,taser,wormlog}.
In this work, we consider \emph{system entities} as \emph{files}, \emph{processes}, and \emph{network connections} in our data model.
We define an interaction among entities as an \emph{event}, which is represented using the triple \emph{$\langle$subject, operation, object$\rangle$}.
We categorize events into three types according to the type of their object entities, namely \emph{file events}, \emph{process events}, and \emph{network connection events}.

Entities and events have various attributes (\cref{tab:entity-attributes,tab:event-attributes}).
The attributes of an entity include the properties to describe the entities (\eg file name, process name, and IP addresses),
and the unique identifiers to distinguish entities (\eg file data ID and process ID).
The attributes of an event include event origins (\ie agent ID and start time/end time),
operations (\eg file read/write),
and other security-related properties (\eg failure code).
In particular, agent ID refers to the unique ID of the host where the entity/event is collected.

\begin{table}[!t]
	\centering
	\caption{Representative attributes of system entities}\label{tab:entity-attributes}
	\begin{adjustbox}{width=0.48\textwidth}
		\begin{tabular}{|l|l|}
			\hline
			\textbf{Entity}		&\textbf{Attributes}\\\hline
			File				&Name, Owner/Group, VolumeID, DataID, etc.\\\hline
			Process			&PID, Name, User, Cmd, Binary Signature, etc.\\\hline
			Network Connection	& IP, Port, Protocol \\\hline
		\end{tabular}
	\end{adjustbox}

	\vspace*{1ex}
\end{table}

\begin{table}[!t]
	\centering
	\caption{Representative attributes of system events}\label{tab:event-attributes}
	\begin{adjustbox}{width=0.48\textwidth}
		\begin{tabular}{|l|l|}
			\hline
			Operation		& Read/Write, Execute, Start/End, Rename/Delete.\\\hline
			Time/Sequence		& Start Time/End Time, Event Sequence\\\hline
			Misc.		& Subject ID, Object ID, Failure Code\\\hline
		\end{tabular}
	\end{adjustbox}

		\vspace*{1ex}
\end{table}

\subsection{\dsl Queries for Anomalies}
\label{subsec:saqlquery}
We next present how to use \dsl as a unified interface to specify various types of abnormal system behaviors.

\para{Rule-based Anomaly}
Advanced cyber attacks typically include a series of steps that exploit vulnerabilities across multiple systems for stealing sensitive information~\cite{aptfireeye,aptsymantec}.
\cref{query:rule} shows a \dsl query for describing an attack step that reads external network (\incode{evt1}), downloads a database cracking tool {\tt gsecdump.exe} (\incode{evt2}), and executes (\incode{evt3}) it to obtain database credentials.
It also specifies these events should occur in ascending temporal order (Line 4).

\begin{lstlisting}[captionpos=b, caption={A rule-based \dsl query}, label={query:rule}]
proc p1 read || write ip i1[src_ip != "internal_address"] as evt1
proc p2["%powershell.exe"] write file f1["%gsecdump.exe"] as evt2
proc p3["%cmd.exe"] start proc p4["%gsecdump.exe"] as evt3
with evt1 -> evt2 -> evt3
return p1, i1, p2, f1, p3, p4 // p1 -> p1.exe_name, i1 -> i1.dst_ip, f1 -> f1.name
\end{lstlisting}

\para{Time-Series Anomaly}
\dsl query provides the constructs of sliding windows to enable the specification of time-series anomaly models.
For example, a \dsl query may monitor the amount of data sent out by certain processes and detect unexpectedly large amount of data transferred within a short period. This type of query can detect network spikes~\cite{netspike,splunkspike},
which often indicates a data exfiltration.
\cref{query:frequency} shows a \dsl query that monitors network usage of each application and raises an alert when the network usage is abnormally high.
It specifies a 10-minute sliding window (Line 1), collects the amount of data sent through network within each window (Lines 2-4),
and computes the moving average to detect spikes of network data transfers (Line 5). 
In the query, \incode{ss[0]} means the state of the current window while \incode{ss[1]} and \incode{ss[2]} represent the states of the two past windows respectively (\incode{ss[2]} occurs earlier than \incode{ss[1]}).
Existing stream query systems and anomaly systems~\cite{cql, streamanomaly1,MillWheel} lack the expressiveness of stateful computation in sliding windows
to support such anomaly models.

\begin{lstlisting}[captionpos=b, caption={A time-series \dsl query}, label={query:frequency}]
proc p write ip i as evt #time(10 min)
state[3] ss {
	avg_amount := avg(evt.amount)
} group by p
alert (ss[0].avg_amount > (ss[0].avg_amount + ss[1].avg_amount + ss[2].avg_amount) / 3) && (ss[0].avg_amount > 10000)
return p, ss[0].avg_amount, ss[1].avg_amount, ss[2].avg_amount
\end{lstlisting}

\begin{figure*}
	\centering
	\includegraphics[width=0.9\textwidth]{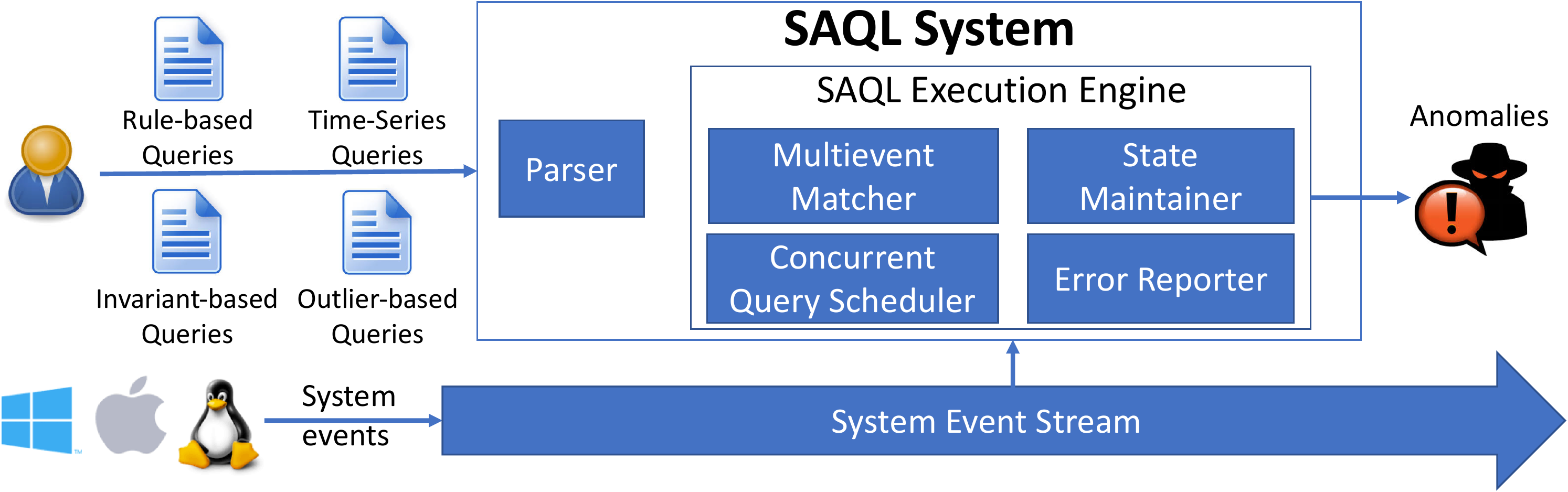}
	\caption{The architecture of \dsl system}
	\label{fig:arch}
\end{figure*}

\para{Invariant-based Anomaly}
Invariant-based anomalies capture the invariants during training periods as models, and use the models later to detect anomalies.
To achieve invariant-based anomaly detection, \dsl provides constructs of invariant models and learning specifics to define and learn invariants of system behaviors,
which allows users to combine both stateful computation and invariants learned under normal operations to detect more types of abnormal system behaviors~\cite{onlinelearning}.
\cref{query:invariant} shows a \dsl query that specifies a 10-second sliding window (Line 1), maintains a set of child processes spawned by the Apache process (Lines 2-4), uses the first ten time windows for training the model (Lines 5-8), and starts to detect abnormal child processes spawned by the Apache process (Line 10).
The model specified in the Lines 5-8 is the set of names of the processes forked by the Apache process in the training stage.
During the online detection phase, this query generates alerts when a process with a new name is forked by the Apache process.
General stream query systems without the support of stateful computation and invariant models cannot express such types of anomaly models.
Note that the invariant definition allows multiple aggregates to be defined.

\begin{lstlisting}[captionpos=b, caption={An invariant-based \dsl query}, label={query:invariant}]
proc p1["%apache.exe"] start proc p2 as evt #time(10 s)
state ss {
	set_proc := set(p2.exe_name)
} group by p1
invariant[10][offline] {
	a := empty_set // invariant init
	a = a union ss.set_proc //invariant update
}
alert |ss.set_proc diff a| > 0
return p1, ss.set_proc
\end{lstlisting}

\para{Outlier-based Anomaly}
Outlier-based anomalies allow users to identify abnormal system behavior through peer comparison, \eg finding outlier processes by comparing the abnormal processes with other peer processes.
To detect outlier-based anomalies, \dsl provides constructs of outlier models to define which information in a time window forms a multidimensional point and compute clusters to identify outliers.
\cref{query:cluster} shows a \dsl query that (1) specifies a 10-minute sliding window (Line 2), (2) computes the amount of data sent through network by the \incode{sqlservr.exe} process for each outgoing IP address (Lines 3-5), and (3) identifies the outliers using DBSCAN clustering (Lines 6-8) to detect the suspicious IP that triggers the database dump.
Note that Line 6 specifies which information of the state forms a point and how the ``distance'' among these points should be computed (``\incode{ed}'' representing Euclidean Distance).
These language constructs enable \dsl to express models for peer comparison, which has limited support from the existing querying systems where only simple aggregation such as max/min are supported~\cite{cql,siddhi,esper}.

\begin{lstlisting}[captionpos=b, caption={An outlier-based \dsl query using clustering}, label={query:cluster}]
agentid = 1 // sqlserver host
proc p["%sqlservr.exe"] read || write ip i as evt #time(10 min)
state ss {
	amt := sum(evt.amount)
} group by i.dstip
cluster(points=all(ss.amt), distance="ed", method="DBSCAN(100000, 5)")
alert cluster.outlier && ss.amt > 1000000
return i.dstip, ss.amt
\end{lstlisting}

In addition to querying outliers through clustering, \dsl also supports querying through aggregation comparison. For example, in \cref{query:cluster}, replacing the \incode{alert} statement with \incode{alert ss.amt>1.5*iqr(all(ss.amt))+q3(all(ss.amt))} gives interquartile range (IQR)-based outlier detection~\cite{casella2002statistical}, and replacing the \incode{alert} statement with \incode{alert ss.amt>3*stddev(all(ss.amt))+avg(all(ss.amt))} gives 3-sigma-based outlier detection~\cite{casella2002statistical}.
\dsl also supports querying outliers through sorting, and reports top sorted results as alerts, which is useful in querying most active processes or IP addresses.


\section{System Overview and Threat Model}
\label{sec:overview}

 \cref{fig:arch} shows the \dsl system architecture.
 We deploy monitoring agents across servers, desktops and laptops in the enterprise to monitor system-level activities by collecting information about system calls from kernels. System monitoring data for Windows, Linux, and Mac OS are collected via ETW event tracing~\cite{etw}, Linux Audit Framework~\cite{auditd}, and DTrace~\cite{dtrace}. The collected data is 
sent to the central server, forming an event stream.
  
  The \dsl system takes \dsl queries from users, and reports the detected alerts over the event stream. The system consists of two 
components: 
  (1) the language parser, implemented using ANTLR 4~\cite{antlr}, performs syntactic and semantic analysis of the input queries and generates an anomaly model context for each query. 
  An anomaly model context is an object abstraction of the input query that contains all the required information for the query execution and anomaly detection;
  (2) the execution engine, built upon Siddhi~\cite{siddhi},  monitors the data stream and reports the detected alerts based on the execution of the anomaly model contexts. 

  The execution engine has four sub-modules:
  (1) the multievent matcher matches the events in the stream against the event patterns specified in the query;
  (2) the state maintainer maintains the states of each sliding window computed from the matched events;  
  (3) the concurrent query scheduler divides the concurrent queries into groups based on the master-dependent-query scheme (\cref{subsec:optimization}) to minimize the need for data copies;
  (4) the error reporter reports errors during the 
  execution.

\para{Threat Model}
\dsl is a stream-based query system over system monitoring data,
and thus we follow the threat model of previous works on system monitoring data~\cite{backtracking,backtracking2,loggc, lee2013high,trustkernel,aiql}. 
We assume that the system monitoring data collected from kernel space~\cite{auditd,etw} are not tampered, and that the kernel is trusted.
Any kernel-level attack that deliberately compromises security auditing systems is beyond the scope of this work.

We do consider that insiders or external attackers have full knowledge of the deployed \dsl queries and the anomaly models. 
They can launch attacks with seemingly ``normal'' activities to evade \dsl's anomaly detection,
and may hide their attacks by mimicking peer hosts' behaviors to avoid \dsl's outlier detection.

\section{\dsl Language Design}
\label{sec:language}
\dsl is designed to facilitate the task of expressing anomalies based on the domain knowledge of experts.
\dsl provides explicit constructs to specify system entities/events, as well as event relationships. 
This facilitates the specification of rule-based anomalies to detect known attack behaviors or enforce enterprise-wide security policies.
\dsl also provides constructs for sliding windows and stateful computation that allow stateful anomaly models to be computed in each sliding window over the data stream. 
This facilitates the specification of time-series anomalies, invariant-based anomalies, and outlier-based anomalies, which lack support from existing stream query systems and stream-based anomaly detection systems.
\cref{bnf:parser} shows the representative rules of \dsl. 
We omit the terminal symbols.


\begin{BNF}[!h]
\footnotesize

\begin{mdframed}

\setlength{\grammarparsep}{-2pt} 
\setlength{\grammarindent}{8em} 

\begin{grammar}
<saql> ::= (<global_cstr>)* (<evt_patt>)+ <temp_rel>? <state>? <groupby>? <alert>? <return> <sortby>? <top>?
\end{grammar}

\textbf{Data types:}
\begin{grammar}
<num> ::= <int> | <float>

<val> ::= <int> | <float> | <string>

<val_set> ::= `('  <val> (`,' <val>)* `)'

<id> ::= <letter>(<letter> | <digit>)*

<attr> ::= <id> (`[' <int> `]')? (`.' <id>)?
\end{grammar}

\vspace*{1ex}

\textbf{Multievent pattern matching:}
\begin{grammar}
<global_cstr> ::= <attr_exp> 

<evt_patt> ::= <entity> <op_exp> <entity> <evt>? <wind>?

<entity> ::= <entity_type> <id> (`[' <attr_exp>`]')?

<op_exp> ::= <op>
\alt `!'<op_exp>
\alt <op_exp> (`&&' | `||') <op_exp>
\alt `(' <op_exp> `)'

<evt> ::= `as' <id> (`[' <attr_exp>`]')?

<wind> ::= `#' <time_wind> | <length_wind>

<time_wind> ::= `time' `(' <num> <time_unit>`)'  (`['<num> <time_unit>`]')?

<length_wind> ::= `length' `(' <int>`)'

<attr_exp> ::=  <attr> | <val>
\alt <attr_exp> <bop> <attr_exp>
\alt <attr_exp> (`&&' | `||') <attr_exp>
\alt `!'<attr_exp> 
\alt `(' <attr_exp> `)'
\alt <attr> `not'? `in' <val_set> 
\alt <agg_func> `(' <attr_exp> (`,' <attr_exp>)*`)' 
\alt <attr_exp> <set_op> <attr_exp> 
\alt `|' <attr_exp> `|' 
\alt <peer_ref> `(' <attr_exp>`)' 


<temp_rel> ::= `with' <id> ((`->'|`<-') (`[' <num> `-' <num> <time_unit>`]')? <id>)+ 

\end{grammar}

\vspace*{1ex}

\textbf{Stateful computation:}
\begin{grammar}
<state> ::= <state_def> <state_inv>? <state_cluster>? 

<state_def> ::= `state' (`[' <int> `]')? <id> `{' <state_field> <state_field>*`}' <groupby>

<state_field> ::= <id> `:=' (<agg_func> | <set_func>) `(' <attr> `)' <groupby>?

<state_inv> ::= `invariant' `[' <int> `]' `[' <train_type> `]'? `{' <inv_init>+ <inv_update>+ `}'

<inv_init> ::= <id> `:=' (<num>|<empty_set>)

<inv_update> ::= <id> `=' <attr_exp>

<state_cluster> ::= `cluster'  `(' <point_def> `,' <distance_def> `,' <method_def> `)'

<point_def> ::= `points' `=' <peer_ref> `(' <attr> (`,' <attr>)* `)' 

<distance_def> ::= `distance' `=' <dist_metric> 

<method_def> ::= `method' `=' <cluster_method> `(' <num> (`,' <num>)* `)'
\end{grammar}

\vspace*{1ex}

\textbf{Alert condition checking:}
\begin{grammar}
<alert> ::= `alert' <attr_exp>
\end{grammar}

\vspace*{1ex}

\textbf{Return and filters:}
\begin{grammar}
<return> ::= `return' <res_pair> (`, ' <res_pair>)*

<res_pair> ::= <attr_exp> (`as' <id>)?

<groupby> ::= `group by' <attr> (`,' <attr>)*

<sortby> ::= `sort by' <attr> (`,' <attr>)* (`asc' | `desc')?

<top> ::= `top' <int>
\end{grammar}

\end{mdframed}

\caption{Representative BNF grammar of \dsl}\label{bnf:parser}
\end{BNF}


\subsection{Multievent Pattern Matching}
\label{subsec:multievent}

\dsl provides the event pattern syntax (in the format of \emph{\{subject-operation-object\}}) to describe system activities, where system entities are represented as subjects and objects, and interactions are represented as operations initiated by subjects and targeted on objects. Besides, the syntax directly supports the specification of event temporal relationships and attribute relationships, which facilitates the specification of complex system behavioral rules.

\para{Global Constraint}
The \emph{$\langle$global_cstr$\rangle$} rule specifies the constraints for all event patterns (\eg \incode{agentid = 1} in \cref{query:cluster} specifies that all event patterns occur on the same host).

\para{Event Pattern}
The \emph{$\langle$evt_patt$\rangle$} rule specifies an event pattern, including the subject/object entity (\emph{$\langle$entity$\rangle$}), the event operation (\emph{$\langle$op_exp$\rangle$}), the event ID (\emph{$\langle$evt$\rangle$}), and the optional sliding window (\emph{$\langle$wind$\rangle$}). 
The \emph{$\langle$entity$\rangle$} rule consists of the entity type (file, process, network connection), the optional entity ID, and the optional attribute constraints expression (\emph{$\langle$attr_exp$\rangle$}). 
Logical operators ($\&\&$, $||$, $!$) can be used in \emph{$\langle$op_exp$\rangle$} to form complex operation expressions (\eg \incode{proc p read || write file f}). 
The \emph{$\langle$attr_exp$\rangle$} rule specifies an attribute expression which supports the use of the logical operators, the comparison operators ($=$, $!=$, $>$, $>=$, $<$, $<=$), the arithmetic operators ($+$, $-$, $*$, $/$), the aggregation functions, and the stateful computation-related operators (\eg \incode{proc p[pid = 1 && name = "\%chrome.exe"]}).

\para{Sliding Window}
The \emph{$\langle$wind$\rangle$} rule specifies the sliding windows for stateful computation. 
For example, \incode{#time(10 min)} in \cref{query:frequency} specifies a sliding window whose width is 10 minutes.
An optional step size can be provided (\eg \incode{#time(10 min)(1 min)} indicates a step size of 1 minute).

\para{Event Temporal Relationship}
The \emph{$\langle$temp_rel$\rangle$} rule specifies the temporal dependencies among event patterns.
For example, \incode{evt1->evt2->evt3} in \cref{query:rule} specifies that \incode{evt1} occurs first, then \incode{evt2}, and finally \incode{evt3}. 
Finer-grained control of temporal distance can also be provided.
For example, \incode{evt1 ->[1-2 min] evt2 ->[1-2 min] evt3} indicates that the time span between the two events is 1 to 2 minutes.

\para{Event Attribute Relationship}
Event attribute relationships can be included in the alert rule (\emph{$\langle$alert$\rangle$}) to specify the attribute dependency of event patterns (\eg  \incode{alert evt1.agentid = evt2.agentid && evt1.dst_id = evt2.src_id} for two event patterns \incode{evt1} and \incode{evt2} indicates that the two events occur at the same host and are ``physically connected'': the object entity of \incode{evt1} is exactly the subject entity of \incode{evt2}).

\para{Context-Aware Syntax Shortcuts}
\begin{itemize}[noitemsep, topsep=1pt, partopsep=1pt, listparindent=\parindent, leftmargin=*]
	\item \emph{Attribute inferences}: 
	(1) default attribute names will be inferred if only attribute values are specified in an event pattern, or only entity IDs are specified in event return. 
	We select the most commonly used attributes in security analysis as default attributes: \incode{name} for files, \incode{exe_name} for processes, and \incode{dst_ip} for network connections.
	For example, in \cref{query:rule}, \incode{file f1["\%gsecdump.exe"]} is equivalent to \incode{file f1[name="\%gsecdump.exe"]}, and \incode{return p1} is equivalent to \incode{return p1.exe_name};
	(2) \incode{id} will be used as default attribute if only entity IDs are specified in the alert condition. 
	For example, given two processes \incode{p1} and \incode{p2}, \incode{alert p1 = p2} is equivalent to \incode{alert p1.id = p2.id}.
	
	\item \emph{Optional ID}: the ID of entity/event can be omitted if it is not referenced in event relationships or event return. 
	For example, in \incode{proc p open file}, we can omit the file entity ID if we will not reference its attributes later.
	
	\item \emph{Entity ID Reuse}: Reused entity IDs in multiple event patterns implicitly indicate the same entity.
\end{itemize}


\subsection{Stateful Computation}
\label{subsec:state}

Based on the constructs of sliding windows, \dsl provides constructs for stateful computation, which consists of two major parts:
defining states based on sliding windows and accessing states of current and past windows to specify time-series anomalies, invariant-based anomalies, and outlier-based anomalies.

\para{State Block}
The \emph{$\langle$state_def$\rangle$} rule specifies a state block by specifying the state count, block ID, and multiple state fields. 
The state count indicates the number of states for the previous sliding windows to be stored (\eg Line 2 in \cref{query:frequency}). 
If not specified, only the state of the current window is stored by default (\eg Line 2 in \cref{query:invariant}).
The \emph{$\langle$state_field$\rangle$} rule
specifies the computation that needs to be performed over the data in the sliding window, and associates the computed value with a variable ID. \dsl supports a broad set of numerical aggregation functions (\eg \incode{sum}, \incode{avg}, \incode{count}, \incode{median}, \incode{percentile}, \incode{stddev}, etc.) and set aggregation functions (\eg \incode{set}, \incode{multiset}). 
After specifying the state block, security analysts can then reference the state fields via the state ID to construct time-series anomaly models (\eg Line 5 in \cref{query:frequency} specifies a three-period simple moving average (SMA)~\cite{hamilton1994time} time-series model to detect network spikes).


\para{State Invariant}
The \emph{$\langle$state_inv$\rangle$} rule specifies invariants of system behaviors and updates these invariants using states computed from sliding windows (i.e., invariant training), so that users can combine both states of windows and invariants learned to detect more types of abnormal system behaviors. 
For example, Lines 5-8 in \cref{query:invariant} specifies an invariant \incode{a} and trains it using the first 10 window results.

\para{State Cluster}
The \emph{$\langle$state_cluster$\rangle$} rule specifies clusters of system behaviors, so that users can identify abnormal behaviors through peer comparison. The cluster specification requires the specification of the points using peer reference keywords $\langle$peer_ref$\rangle$ (\eg \incode{all}), distance metric, and clustering method. \dsl supports common distance metrics (\eg Manhattan distance, Euclidean distance) and major clustering algorithms (\eg K-means~\cite{han2011data}, DBSCAN~\cite{ester1996density}, and hierarchical clustering~\cite{han2011data}). 
For example, Line 6 in \cref{query:cluster} specifies a cluster of the one-dimensional points \incode{ss.amt} using Euclidean distance and DBSCAN algorithm.
\dsl also provides language extensibility that allows other clustering algorithms and metrics to be used through mechanisms such as Java Native Interface (JNI) and Java Naming and Directory Interface (JNDI).

\subsection{Alert Condition Checking}
\label{subsec:alert}
The \emph{$\langle$alert$\rangle$} rule specifies the condition (a boolean expression) for triggering the alert. This enables \dsl to specify a broad set of detection logics for time-series anomalies (\eg Line 5 in \cref{query:frequency}), invariant-based anomalies (\eg Line 9 in \cref{query:invariant}), and outlier-based anomalies (\eg Line 7 in \cref{query:cluster}). Note that in addition to the moving average detection logic specified in \cref{query:frequency}, the flexibility of \dsl also enables the specification of other well-known logics, such as 3-sigma rule~\cite{casella2002statistical} (\eg \incode{alert ss.amt>3*stddev(all(ss.amt))+avg(all(ss.amt))}) and 
IQR rule~\cite{casella2002statistical} (\eg \incode{alert ss.amt>1.5*iqr(all(ss.amt))+q3(all(ss.amt))}).

\subsection{Return and Filters}
\label{subsec:return}
The \emph{$\langle$report$\rangle$} rule specifies the desired attributes of the qualified events to return as results. Constructs such as \incode{group by}, \incode{sort by}, and \incode{top} can be used for further result manipulation and filtering. 
These constructs are useful for querying the most active processes and IP addresses, as well as specifying threshold-based anomaly models without explicitly defining states. 
For example, \cref{query:simple-threshold} computes the IP frequency of each process in a 1-minute sliding window and returns the active processes with a frequency greater than 100.

\begin{lstlisting}[captionpos=b, caption={Threshold-based IP Frequency Anomaly}, label={query:simple-threshold}]
proc p start ip i as evt #time(1 min)
group by p
alert freq > 100
return p, count(i) as freq
\end{lstlisting}

\section{\dsl Execution Engine}
\label{sec:engine}

The \dsl execution engine in \cref{fig:arch} takes the event stream as input, executes the anomaly model contexts generated by the parser, and reports the detected alerts.
To make the system more scalable in supporting multiple concurrent queries, the 
engine employs a \emph{master-dependent-query scheme} that groups semantically compatible queries to share a single copy of the stream data for query execution. 
In this way, the \dsl system significantly reduces the data copies of the stream.


\subsection{Query Execution Pipeline}
\label{subsec:pipeline}
The query engine is built upon Siddhi~\cite{siddhi}, so that our \dsl can leverage its mature stream management engine in terms of event model, stream processing, and stream query.
Given a \dsl query, the parser performs syntactic analysis and semantic analysis to generate an anomaly model context. 
The concurrent query scheduler inside the query optimizer analyzes the newly arrived anomaly model context against the existing anomaly model contexts of the queries that are currently running, and computes an optimized execution schedule by leveraging the master-dependent-query scheme.
The multievent solver analyzes event patterns and their dependencies in the \dsl query, and retrieves the matched events by issuing a Siddhi query to access the data from the stream. 
If the query involves stateful computation, the state maintainer leverages the intermediate execution results to compute and maintain query states. 
Alerts will be generated if the alert conditions are met for the queries.

\subsection{Concurrent Query Scheduler}
\label{subsec:optimization}

\begin{algorithm}[!h]
\footnotesize

\SetAlgoLined
\KwInput{User submitted new SAQL query: $newQ$\newline
Map of concurrent master-dependent queries: $M = \{masQ_i \rightarrow \{depQ_{ij}\}\}$
}
\KwOutput{Execution results of $newQ$
}


\uIf{$M.isEmpty$}{
		
	\Return $execAsMas(newQ, M)$\;
}
\Else{
	\For{$masQ_i$ in $M.keys$}{
		$covQ = constructSemanticCover(masQ_i, newQ)$\;
		
		\If{$covQ \neq null$}{
			\If{$covQ \neq masQ_i$}{
				$replMas(masQ_i, covQ, M)$\;
			}
			$addDep(covQ, newQ)$\;
			
			\Return $execDep(newQ, covQ)$\;
		}
	}

	\Return $execAsMas(newQ, M)$\;
}
\vspace{1ex}

\Fn{$constructSemanticCover($masQ$, $newQ$)$}{
	
	\If{Both $masQ$ and $newQ$ define a single event pattern}{
		\If{$masQ$ and $newQ$ share the same event type, operation type, and sliding window type}{
			Construct the event pattern cover $evtPattCovQ$ by taking the union of their attributes and agent IDs and the GCD of their window lengths\;
			\If{Both $masQ$ and $depQ$ define states}{
				\If{$masQ$ and $depQ$ have the same sliding window length \textbf{and} $masQ$ defines a super set of state fields of $depQ$}{
					Construct the state cover $stateCovQ$ by taking the union of their state fields\;
				}
			}
			\Return $covQ$ by concatenating $evtPattCovQ$, $stateCovQ$, and the rest parts of $masQ$\;
		}
	}

	\Return null\;
}

\Fn{$execAsMas(newQ, M)$}{
	Make $newQ$ as a new master and execute it\;

}


\Fn{$addDep(masQ, depQ, M)$}{
	Add $depQ$ to the dependencies of $masQ$\;
}

\Fn{$replMas(oldMasQ, newMasQ, M)$}{
	Replace the old master $oldMasQ$ with the new master $newMasQ$ and update dependencies\;
%
}



\Fn{$execDep (depQ, masQ)$}{
	\uIf{$depQ$ == $masQ$}{
		\Return execution results of $masQ$\;
	}
	\uElseIf{Both $masQ$ and $depQ$ define states}{
		\If{$masQ$ and $depQ$ have the same sliding window length \textbf{and} $masQ$ defines a super set of state fields of $depQ$}{
			Fetch the state aggregation results of $masQ$, enforce additional filters, and feed into the execution pipeline of $depQ$\;
		}
	}
	\Else{
		Fetch the matched events of $masQ$, enforce additional filters, and feed into the execution pipeline of $depQ$\;
	}
}

\caption{Master-dependent-query scheme}\label{alg:schedule}
\end{algorithm}

The concurrent query scheduler in \Cref{fig:arch} schedules the execution of concurrent queries. 
A straightforward scheduling strategy is to make copies of the stream data and feed the copies to each query, allowing each query to operate separately. However, system monitoring produces huge amount of daily logs~\cite{loggc,reduction}, and such copy scheme incurs high 
memory usage, which greatly limits the scalability of the system.

\para{Master-Dependent-Query Scheme}
To efficiently support concurrent query execution, the concurrent query scheduler adopts a \emph{master-dependent-query scheme}. 
In the scheme, only master queries have direct access to the data stream, and the execution of the dependent queries depends on the execution of their master queries.
Given that the execution pipeline of a query typically involves four phases (\ie event pattern matching, stateful computation, alert condition checking, and attributes return), the key idea is to maintain a map $M$ from a master query to its dependent queries, and \emph{let the execution of dependent queries share the intermediate execution results of their master query in certain phases}, so that unnecessary data copies of the stream can be significantly reduced. 
\cref{alg:schedule} shows the scheduling algorithm:
\begin{enumerate}[label={\arabic*.}, noitemsep, topsep=1pt, partopsep=1pt, listparindent=\parindent, leftmargin=*]
	\item The scheme first checks if $M$ is empty (i.e., no concurrent running queries). If so, the scheme sets $newQ$ as a master query, stores it in $M$, and executes it.
	\item If $M$ is not empty, the scheme checks $newQ$ against every master query $masQ_i$ for compatibility
	and tries to construct a semantic cover $covQ$.
	If the construction is successful, the scheme then checks whether $covQ$ equals $masQ_i$.
	\item If $covQ$ is different from $masQ_i$, the scheme updates the master query by replacing 
	$masQ_i$
	with 
	$covQ$ and updates all the dependent queries of $masQ_i$ to $covQ$.
	\item The scheme then adds $newQ$ as a new dependent query of $covQ$, and executes $newQ$ based on $covQ$.
	\item Finally, if there are no master queries found to be compatible with $newQ$, the scheme sets $newQ$ as a new master query, stores it in $M$, and executes it.  
\end{enumerate}


Two key steps in \cref{alg:schedule} are $constructSemanticCover()$ and $execDep()$. 
The construction of a semantic cover requires that (1) the $masQ$ and $depQ$ both define a single event pattern and (2) their event types, operation types, and sliding window types must be the same\footnote{We leave the support for multiple event patterns for future work}. 
The scheme then explores the following four \emph{optimization dimensions}: event attributes, agent ID, sliding window, and state aggregation.
Specifically, the scheme first constructs an event pattern cover by taking the union of the two queries' event attributes and agent IDs, and taking the greatest common divisor (GCD) of the window lengths.
It then constructs a state block cover by taking the union of the two queries' state fields (if applicable), and returns the semantic cover by concatenating the event pattern cover, the state block cover, and the rest parts of $masQ$.

The execution of $depQ$ depends on the execution of $masQ$. If two queries are the same, the engine directly uses the execution results of $masQ$ as the execution results of $depQ$.
Otherwise, the engine fetches the \emph{intermediate results} from the execution pipeline of $masQ$ based on the \emph{level of compatibility}. 
The scheme currently enforces the results sharing in two execution phases: event pattern matching and stateful computation:
(1) if both $dep$ and $masQ$ define states and their sliding window lengths are the same, the engine fetches the state aggregate results of $masQ$;
(2) otherwise, the engine fetches the matched events of $masQ$ without its further state aggregate results.
The engine then enforces additional filters and feed the filtered results into the rest of the execution pipeline of $depQ$ for further execution.

\section{Deployment and Evaluation}
\label{sec:eval}

We deployed the \dsl system in 
NEC Labs America comprising 150 hosts
(10 servers, 140 employee stations; generating around 3750 events/s).
To evaluate the expressiveness of \dsl and the \dsl's overall effectiveness and efficiency, 
we first perform a series of attacks based on known exploits in the deployed environment and construct 17 \dsl queries to detect them. 	
We further conduct a pressure test to measure the maximum performance that our system can achieve. 
Finally, we conduct a performance evaluation on a micro-benchmark (64 queries) to evaluate the effectiveness of our query engine in handling concurrent queries.
In total, our evaluations use 1.1TB of real system monitoring data (containing 3.3 billion system events).
All the attack queries are available in Appendix, and all the micro-benchmark queries are available on our \emph{project website~\cite{saql}}.

\subsection{Evaluation Setup}
\label{subsec:setup}

The evaluations are conducted on a server with an Intel(R) Xeon(R) CPU E1650 (2.20GHz, 12 cores) and 128GB of RAM. 
The server continuously receives a stream of system monitoring data collected from the hosts deployed with the data collection agents. 
We developed a web-based client for query submission and deployed the \dsl system on the server for query execution.
To reproduce the attack scenarios for the performance evaluation in \cref{subsec:eval-perf}, we stored the collected data in databases and developed a stream replayer to replay the system monitoring data from the databases.

\subsection{Attack Cases Study}
\label{subsec:case}

We performed four major types of attack behaviors in the deployed environment based on known exploits:
(1) APT attack~\cite{aptfireeye,aptsymantec},
(2) SQL injection attack~\cite{sqlinjection1,sqlinjection2},
(3) Bash shellshock command injection attack~\cite{shellshock}, and
(4) suspicious system behaviors.

\subsubsection{Attack Behaviors}
\label{subsubsec:attacks}

\begin{figure}[t]
	\centering
	\includegraphics[width=0.48\textwidth]{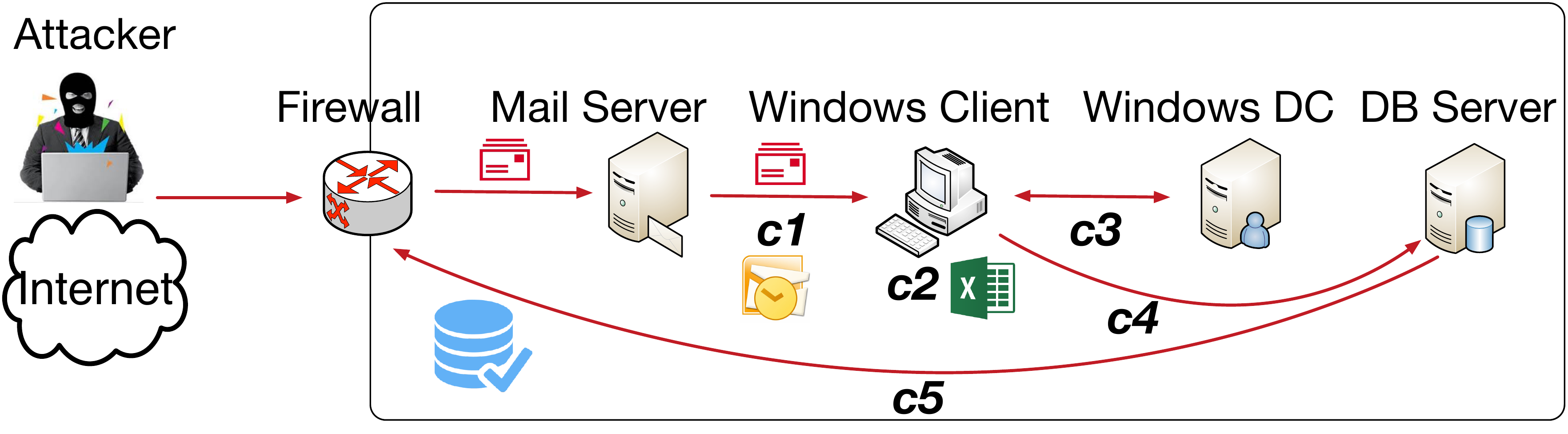}
	\caption{Environmental setup for the APT attack}
	\label{fig:outlook-apt}
\end{figure}

\para{APT Attack}
We ask white hat hackers to perform an APT attack in the deployed environment, as shown in \cref{fig:outlook-apt}. 
Below are the attack steps:

\eat{
Figure~\ref{fig:outlook-apt} shows the setup of the APT attack scenario that is constructed based on existing studies~\cite{aptfireeye,aptsymantec}. 
To mimic the typical enterprise environment, we deployed a virtual system in our deployed enterprise environment, which consists of a gateway router, a mail server, a Windows client, a Windows domain controller, and a SQL database server. 
To be close to the real APT attack where a portion of hosts are compromised by the attacker, we asked a group of white hat hackers to penetrate into the system using several exploits
and steal the valuable information stored in the database server. 
The successful APT attack requires the white hat hackers to perform the following attack steps:
}

\begin{itemize}[label={\arabic*.}, noitemsep, topsep=1pt, partopsep=1pt, listparindent=\parindent, leftmargin=*]
	\item[\emph{c1}] \emph{Initial Compromise}: The attacker sends a crafted email to the victim. The email contains an Excel file with a malicious macro embedded. 
	
	\item[\emph{c2}] \emph{Malware Infection}: The victim opens the Excel file through the Outlook client and runs the macro, which downloads and executes a malicious script (CVE-2008-0081~\cite{cveexcel}) to open a backdoor for the attacker.
	
	\item[\emph{c3}] \emph{Privilege Escalation}: The attacker enters the victim's machine through the backdoor, scans the network ports to discover the IP address of the database, and runs the database cracking tool ({\tt gsecdump.exe}) to steal the credentials of the database.
	
	\item[\emph{c4}] \emph{Penetration into Database Server}: Using the credentials, the attacker penetrates into the database server and delivers a VBScript to drop another malicious script, which creates another backdoor.
	
	\item[\emph{c5}] \emph{Data Exfiltration}: With the access to the database server, the attacker dumps the database content using {\tt osql.exe} and sends the data dump back to his host.
\end{itemize}

For each attack step, we construct a rule-based anomaly query (\ie \cref{query:c1,query:c2,query:c3,query:c4,query:c5}). 
Besides, we construct 3 advanced anomaly queries:

\begin{itemize}[noitemsep, topsep=1pt, partopsep=1pt, listparindent=\parindent, leftmargin=*]
	\item We construct an invariant-based anomaly query (\cref{query:c2:invariant}) to detect the scenario where Excel executes a malicious script that it has never executed before:
	The invariant contains all unique processes started by Excel in the first 100 sliding windows. During the detection phase, new processes that deviate from the invariant will be reported as alerts. 
	This query can be used to detect the unseen suspicious Java process started by Excel (\ie step \emph{c2}).
	
	\item We construct a time-series anomaly query (\cref{query:c5:timeseries}) based on SMA to detect 
the scenario where abnormally high volumes of data are exchanged via network on the database server (\ie step \emph{c5}): 
	For every process on the database server, this query detects the processes that transfer abnormally high volumes of data to the network. 
This query can be used to detect the large amount of data transferred from the database server.
	
	\item We also construct an outlier-based anomaly query (\cref{query:c5:outlier}) to detect processes that transfer high volumes of data to the network (\ie step \emph{c5}): 
	The query detects such processes through peer comparison based on DBSCAN.
	The detection logic here is different from \cref{query:c5:timeseries}, which detects anomalies through comparison with historical states based on SMA.
\end{itemize}

Note that the construction of these 3 queries assumes no knowledge of the detailed attack steps. 

\begin{table*}[!t]
  \centering
    \caption{Execution statistics of 17 \dsl queries for four major types of attacks}
  \label{tab:case}%
 \begin{adjustbox}{width=0.95\textwidth}
    \begin{tabular}{|l|r|r|r|r|r|r|r|}
    \hline
    \textbf{\dsl Query} & \textbf{Alert Detection Latency} & \textbf{Num. of States} & \textbf{Tot. State Size} & \textbf{Avg. State Size} & \textbf{CPU} & \textbf{Memory} \\\hline
    
    \emph{apt-c1} & $\leq$1ms  & N/A   & N/A   & N/A   & 10\%  & 1.7GB \\\hline
    \emph{apt-c2} & $\leq$1ms  & N/A   & N/A   & N/A   & 10\%  & 1.8GB \\\hline
    \emph{apt-c3} & 6ms  & N/A   & N/A   & N/A   & 8\%   & 1.6GB \\\hline
    \emph{apt-c4} & 10ms & N/A   & N/A   & N/A   & 10\%  & 1.5GB \\\hline
    \emph{apt-c5} & 3ms   & N/A   & N/A   & N/A   & 10\%  & 1.6GB \\\hline
    \emph{apt-c2-invariant} & $\leq$1ms & 5	& 5 & 1 & 8\%   & 1.8GB \\\hline
    \emph{apt-c5-timeseries} & $\leq$1ms   & 812	& 3321 & 4.09 & 6\%   & 2.2GB \\\hline
    \emph{apt-c5-outlier} & 2ms   & 812 & 3321 & 4.09 & 8\%   & 2.2GB \\\hline
    \emph{shellshock} & 5ms   & 3 & 3 & 1 & 8\%   & 2.7GB \\\hline
    \emph{sql-injection} & 1776ms & 14 & 13841 & 988.6 & 8\%   & 1.9GB \\\hline
    \emph{dropbox} & 2ms   & N/A   & N/A   & N/A   & 8\%   & 1.2GB \\\hline
    \emph{command-history} & $\leq$1ms & N/A   & N/A   & N/A   & 10\%  & 2.2GB \\\hline
    \emph{password} & $\leq$1ms & N/A   & N/A   & N/A   & 9\%   & 1.6GB \\\hline
    \emph{login-log} & $\leq$1ms & N/A   & N/A   & N/A   & 10\%  & 2.2GB \\\hline
    \emph{sshkey} & $\leq$1ms & N/A   & N/A   & N/A   & 10\%  & 2.1GB \\\hline
    \emph{usb}   & $\leq$1ms & N/A   & N/A   & N/A   & 9\%   & 2.1GB \\\hline
    \emph{ipfreq} & $\leq$1ms & N/A   & N/A   & N/A   & 10\%  & 2.1GB \\\hline
    \end{tabular}%
    \end{adjustbox}
\vspace*{1ex}
\end{table*}%

\eat{
This attack can cause three anomalies. Specifically, these anomalies are:
\begin{enumerate}[label={\arabic*.}, noitemsep, topsep=1pt, partopsep=1pt, listparindent=\parindent, leftmargin=*]
    \item In step \emph{a2}, Excel executes the a program that it has never executed before, which is the malware。
    \item In step \emph{a3}, the execution of {\tt gsecdump.exe}.
    \item In step \emph{a5}, uploading the dumped database can have an abnormally high volume of data exchange with the external Internet. 
\end{enumerate}

In this case, each individual anomaly is not sufficient to identify the attack with very high confidence due to the false positives. For instance, the user can run the {\tt gsecdump.exe} to just recover his password, the user can simply dump the database to backup the data, and finally, the high volume of uploading data  can be caused by a normal  file upload. However, if all three of these anomalies appear together, it is a strong signal for an invasion attack.
}

\para{SQL Injection Attack}
We conduct a SQL injection attack~\cite{halfond06issse} for a typical web application server configuration. 
The setup has multiple web application servers that accept incoming web traffics to load balance.
Each of these web servers connects to a single database server to authenticate users and serves dynamic contents. 
However, these web applications provide limited input sanitization and thus are susceptible to SQL injection attack.

We use SQLMap~\cite{sqlmap} to automate the attack against one of the web application servers. 
In the process of detecting and exploiting SQL injection flaws and taking over the database server, the attack generates an excessive amount of network traffic between the web application server and the database server. 
We construct an outlier-based anomaly query (\cref{query:sql}) to detect abnormally large data transfers to external IP addresses.

\para{Bash Shellshock Command Injection Attack}
We conduct a  command injection attack against a system that installs an outdated Bash package susceptible to the Shellshock vulnerability~\cite{shellshock}. 
With a crafted payload, the attacker initiates a HTTP request to the web server and opens a Shell session over the remote host.  
The behavior of the web server in creating a long-running Shell process is an outlier pattern. 
We construct an invariant-based anomaly query (\cref{query:shellshock}) to learn the invariant of child processes of Apache, and use it to detect any unseen child process (i.e., {\tt /bin/bash} in this attack).

\para{Suspicious System Behaviors}
Besides known threats, security analysts often have their own definitions of suspicious system behaviors, such as accessing credential files using unauthorized software and running forbidden software.
We construct 7 rule-based queries to detect a representative set of suspicious behaviors:

\begin{itemize}[noitemsep, topsep=1pt, partopsep=1pt, listparindent=\parindent, leftmargin=*]
	\item Forbidden Dropbox usage (\cref{query:dropbox}): finding the activities of Dropbox processes.
	\item Command history probing (\cref{query:history}): finding the processes that access multiple command history files in a relatively short period.
	\item Unauthorized password files accesses (\cref{query:password}): finding the unauthorized processes that access the protected password files.
	\item Unauthorized login logs accesses (\cref{query:login}): finding the unauthorized processes that access the log files of login activities.
	\item Unauthorized SSH key files accesses (\cref{query:sshkey}): finding the unauthorized processes that access the SSH key files.
	\item Forbidden USB drives usage (\cref{query:usb}): finding the processes that access the files in the USB drive.
	\item IP frequency analysis (\cref{query:ipfreq}): finding the processes with high frequency network accesses.
\end{itemize}

\eat{
\begin{lstlisting}[captionpos=b, caption={Command history probing}, label={query:command-history}]
proc p read || write file f["%.viminfo"||"%.bash_history"||"%.zsh_history"||"%.lesshst"||"%.pgadmin_histoqueries"||"%.mysql_history"] as evt
report p, f, evt.agentid, evt.starttime, evt.endtime
\end{lstlisting}
\vspace*{-2ex}
}

\subsubsection{Query Execution Statistics}
\label{subsec:case-statistics}

To demonstrate the effectiveness of the \dsl system in supporting timely anomaly detection, we measure the following performance statistics of the query execution:
\begin{itemize}[noitemsep, topsep=1pt, partopsep=1pt, listparindent=\parindent, leftmargin=*]
\item \emph{Alert detection latency}: the difference between the time that the anomaly event gets detected and the time that the anomaly event enters the \dsl engine.
\item \emph{Number of states}: the number of sliding windows encountered from the time that the query gets launched to the time that the anomaly event gets detected.
\item \emph{Average state size}: the average number of aggregation results per state.
\end{itemize}

The results are shown in \cref{tab:case}. 
We observe that:
(1) the alert detection latency is low ($\leq$10ms for most queries and $<$2s for all queries). For \emph{sql-injection}, the latency is a bit larger due to the additional complexity of the specified DBSCAN clustering algorithm in the query;
(2) the system is able to efficiently support 150 enterprise hosts, with $<$ 10\% CPU utilization and $<$2.7GB memory utilization. 
Note that this is far from the full processing power of our system on the deployed server, and our system is able to support a lot more hosts (as experimented in \cref{subsec:pressure});
(3) the number of states and the average state size vary with a number of factors, such as query running time, data volume, and query attributes (\eg number of agents, number of attributes, attribute filtering power). Even though the amount of system monitoring data is huge, a \dsl query often restricts one or several data dimensions by specifying attributes. Thus, the state computation is often maintained in a manageable level.



\subsection{Pressure Test}
\label{subsec:pressure}

\begin{figure}[t]
	\centering
	\includegraphics[width=0.48\textwidth]{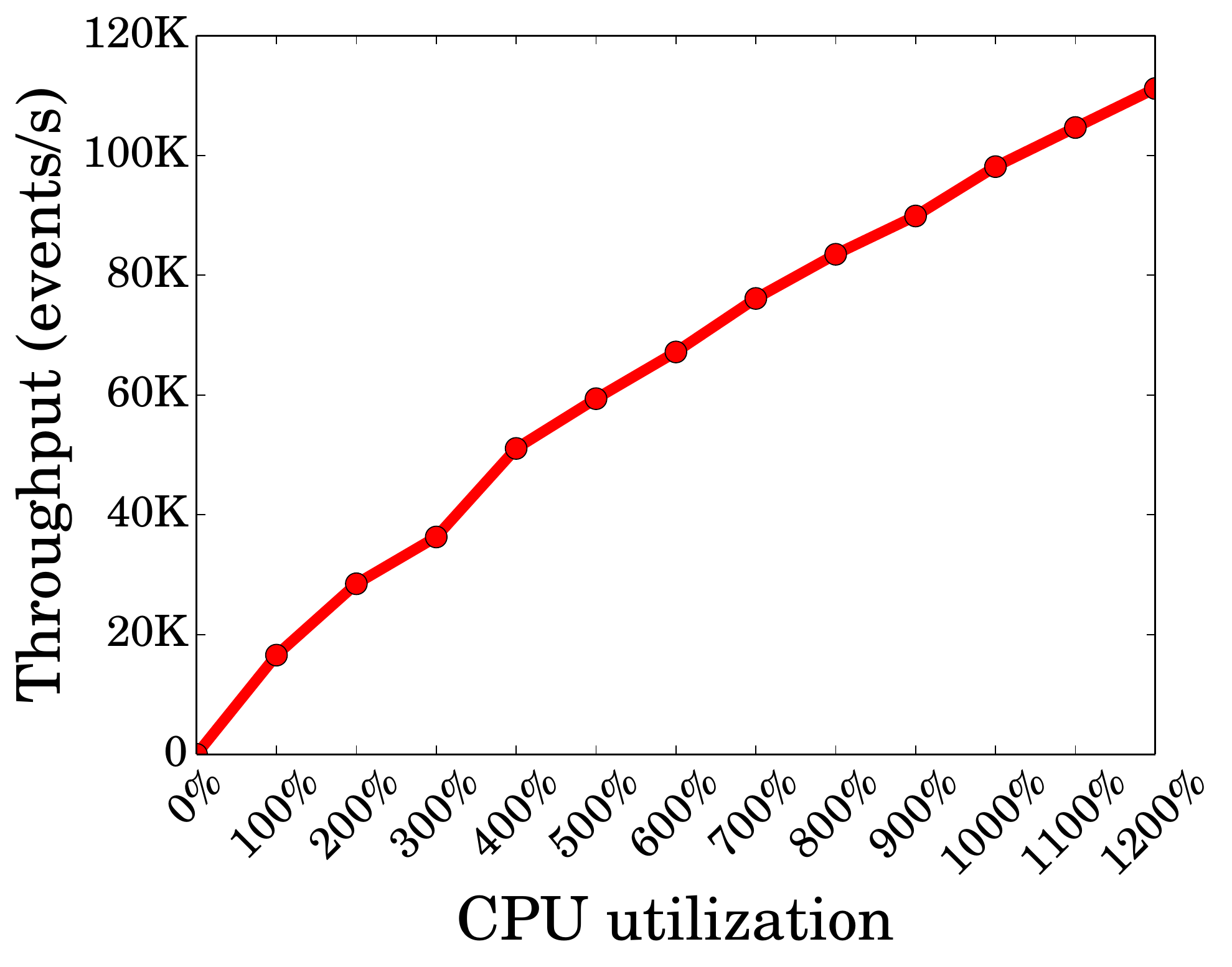}
	\caption{Throughput of the \dsl system under different CPU utilizations. }
	\label{fig:pressure-throughput}
\end{figure}

We conduct a pressure test of our system by replicating the data stream, while restricting the CPU utilization to certain levels~\cite{cpulimit}. 
When we conduct the experiments, we set the maximum Java heap size to be 100GB so that memory will not be a bottleneck.
We deploy a query that retrieves all file events as the representative rule-based query, and measure the system throughput to demonstrate the query processing capabilities of our system.

\para{Evaluation Results}
\cref{fig:pressure-throughput} shows the throughput of the \dsl system under different CPU utilizations. 
We observe that using a deployed server with 12 cores, the \dsl system achieves a maximum throughput of 110000 events/s. 
Given that our deployed enterprise environment comprises 150 hosts with 3750 events generated per second, we can estimate that the \dsl system on this server can support $\sim$4000 hosts.
While such promising results demonstrate that our \dsl system deployed in only one server can easily support far more than hundreds of hosts for many organizations,
there are other factors that can affect the performance of the system.
First, queries that involve temporal dependencies may cause more computation on the query engine, and thus could limit the maximum number of hosts that our \dsl system can support.
Second, if multiple queries are running concurrently, multiple copies of the data stream are created to support the query computation, which would significantly compromise the system performance. 
Our next evaluation demonstrate the impact of concurrent queries and how our master-dependent-query scheme mitigates the problem.


\newcommand{\attFile}{Sensitive file accesses}
\newcommand{\attBrowser}{Browsers access files}
\newcommand{\attNetwork}{Processes access networks}
\newcommand{\attProcess}{Processes spawn}
\newcommand{\evalEvent}{Event attributes}
\newcommand{\evalWindow}{Sliding window}
\newcommand{\evalAgent}{Agent ID}
\newcommand{\evalState}{State aggregation}

\begin{figure*}[!h]
\center
\begin{subfigure}[H]{0.25\textwidth}
  \includegraphics[width=\linewidth]{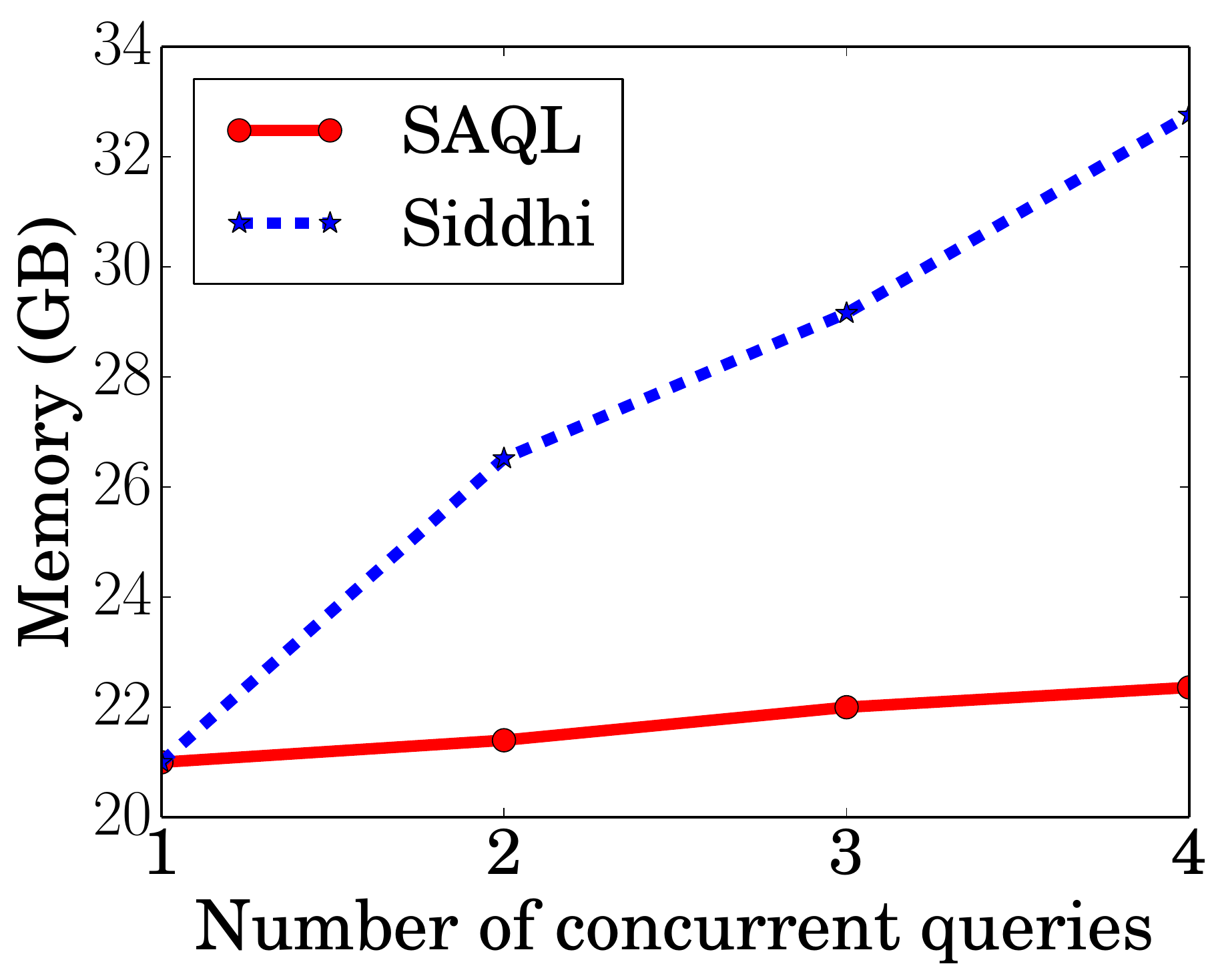}
  \caption{\attFile}
\end{subfigure}%
\hfill
\begin{subfigure}[H]{0.25\textwidth}
  \includegraphics[width=\linewidth]{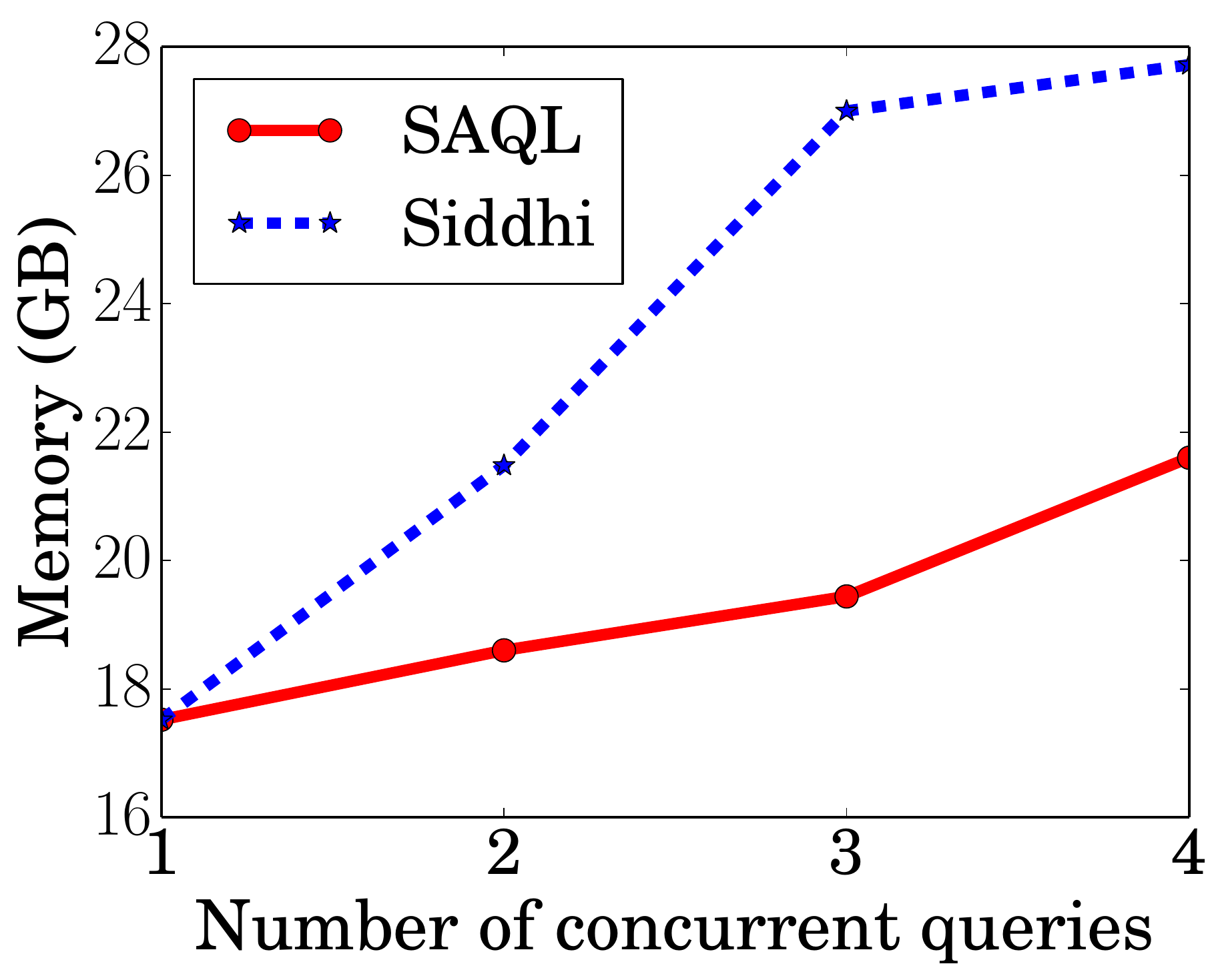}
  \caption{\attBrowser}
\end{subfigure}%
\hfill
\begin{subfigure}[H]{0.25\textwidth}
  \includegraphics[width=\linewidth]{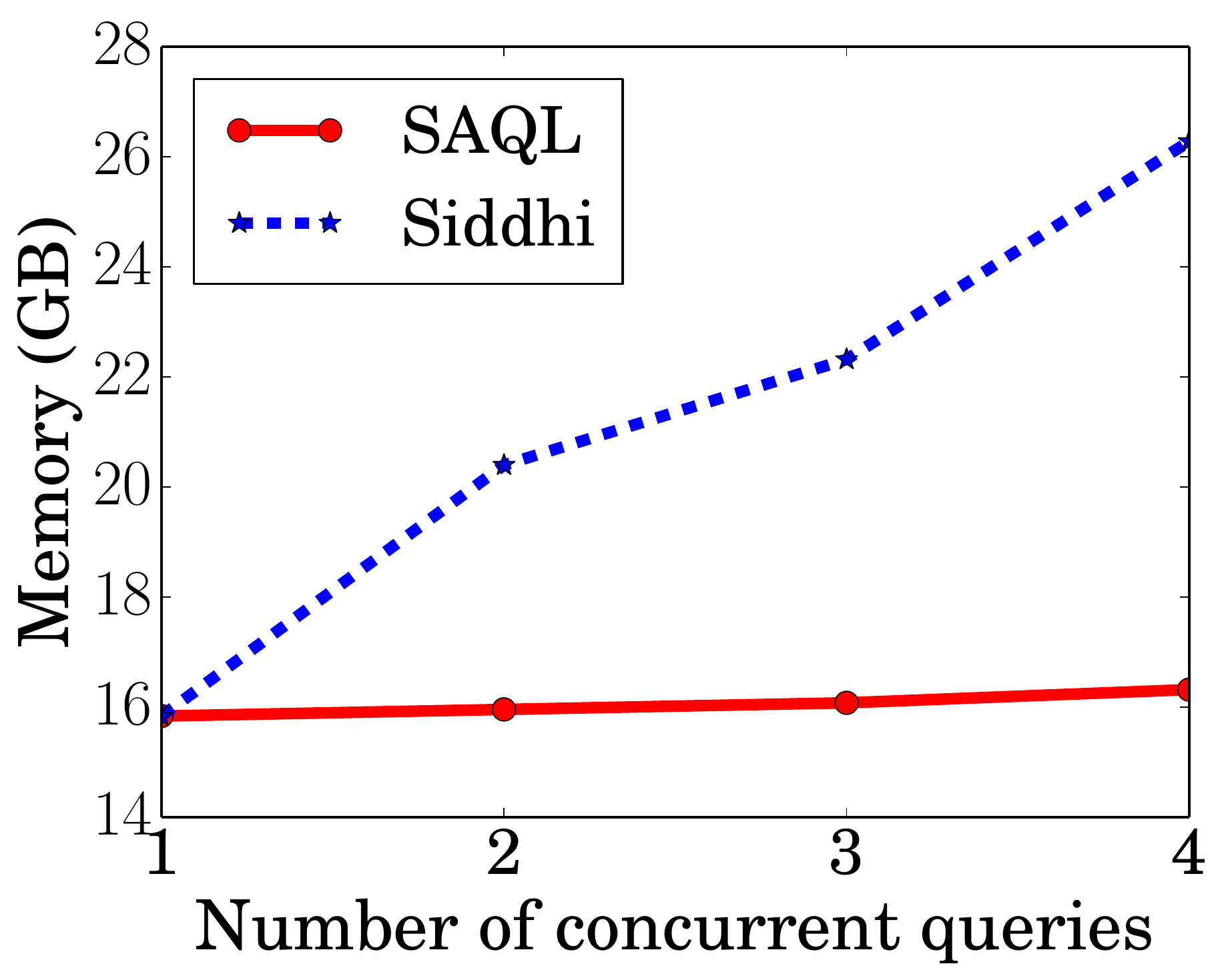}
  \caption{\attNetwork}
\end{subfigure}%
\hfill
\begin{subfigure}[H]{0.25\textwidth}
  \includegraphics[width=\linewidth]{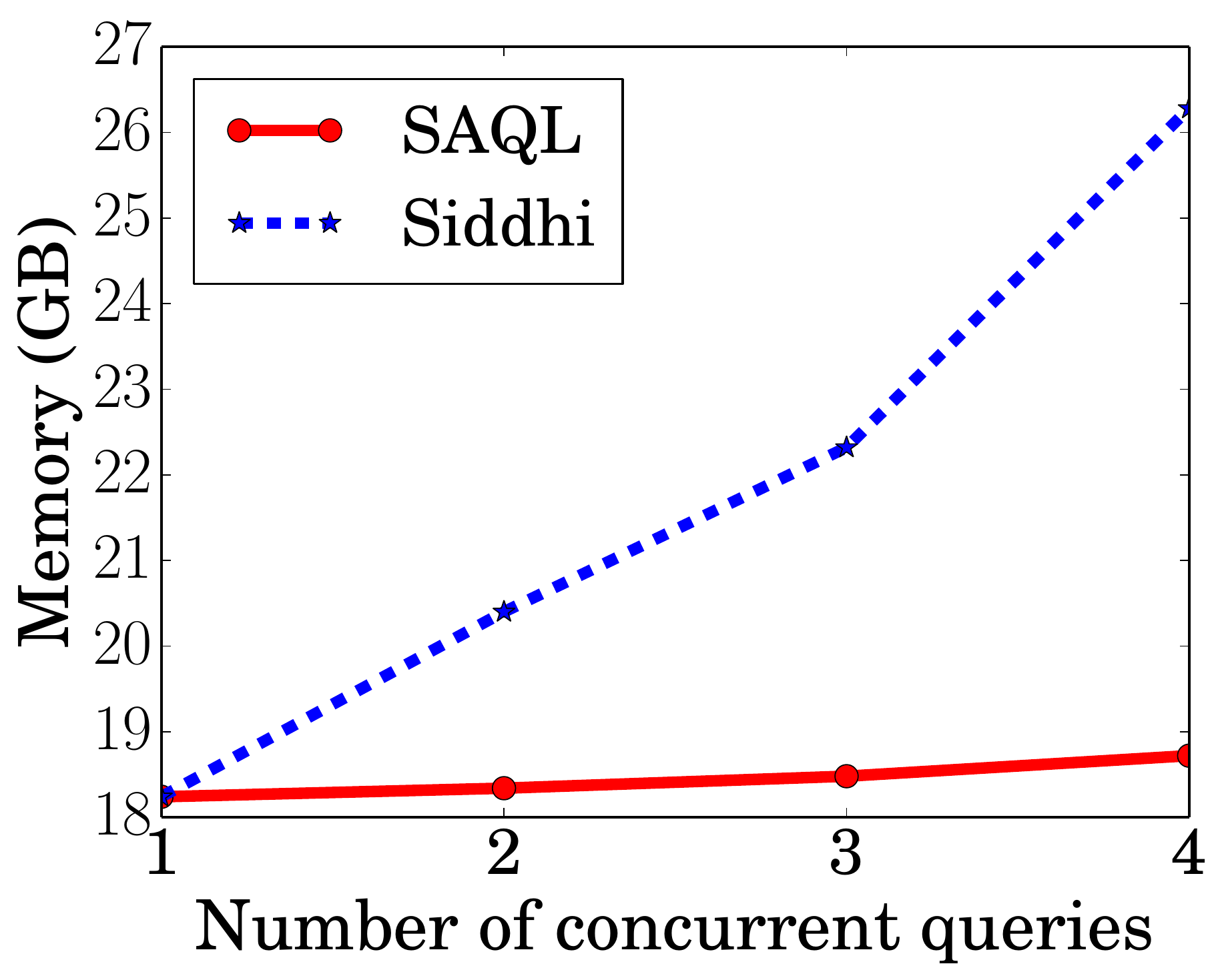}
  \caption{\attProcess}
\end{subfigure}
\caption{\evalEvent}
\label{fig:micro-benchmarks:attribute}
\vspace*{1ex}
\end{figure*}

\begin{figure*}[!h]
\center
\begin{subfigure}[H]{0.25\textwidth}
  \includegraphics[width=\linewidth]{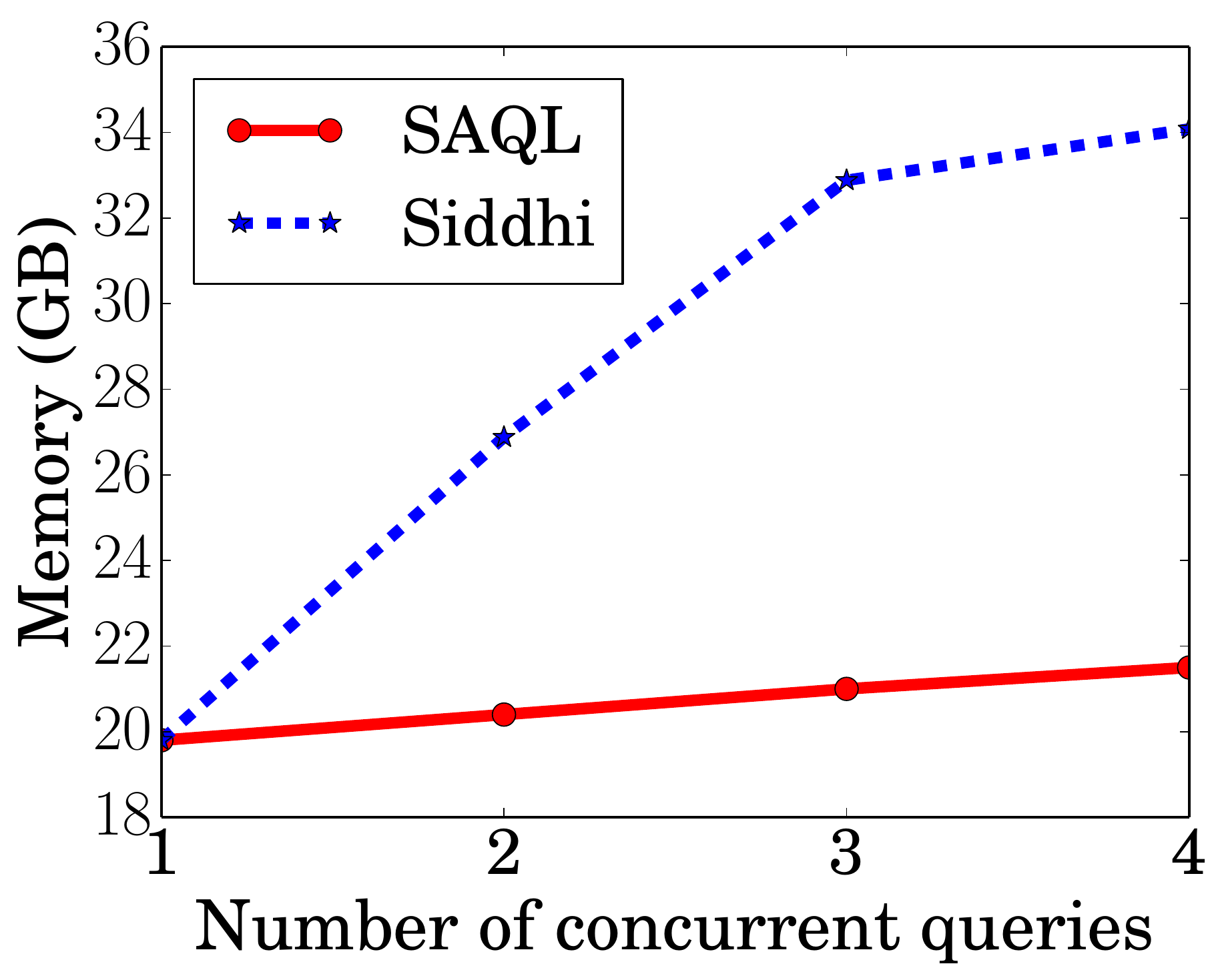}
  \caption{\attFile}
\end{subfigure}%
\hfill
\begin{subfigure}[H]{0.25\textwidth}
  \includegraphics[width=\linewidth]{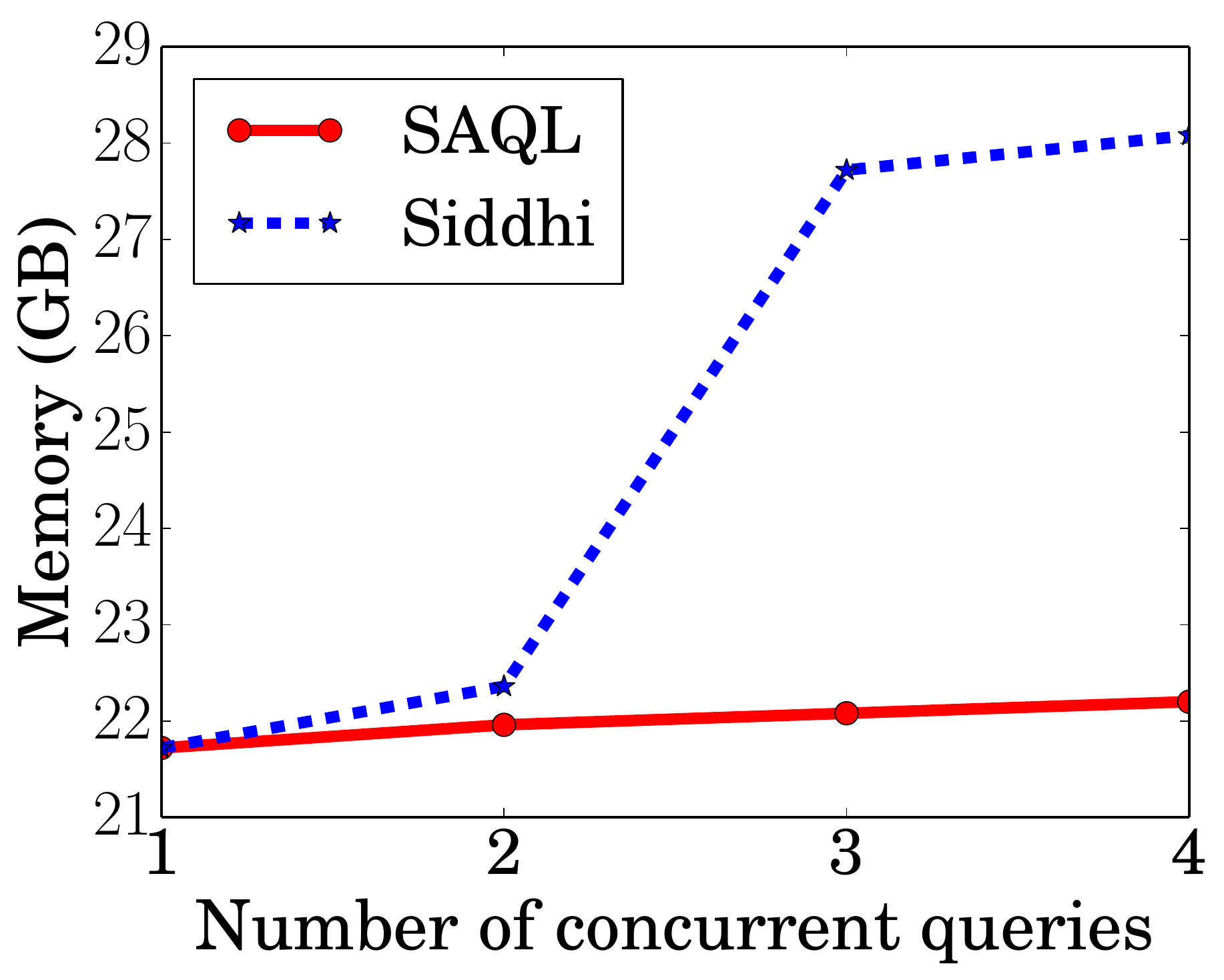}
  \caption{\attBrowser}
\end{subfigure}%
\hfill
\begin{subfigure}[H]{0.25\textwidth}
  \includegraphics[width=\linewidth]{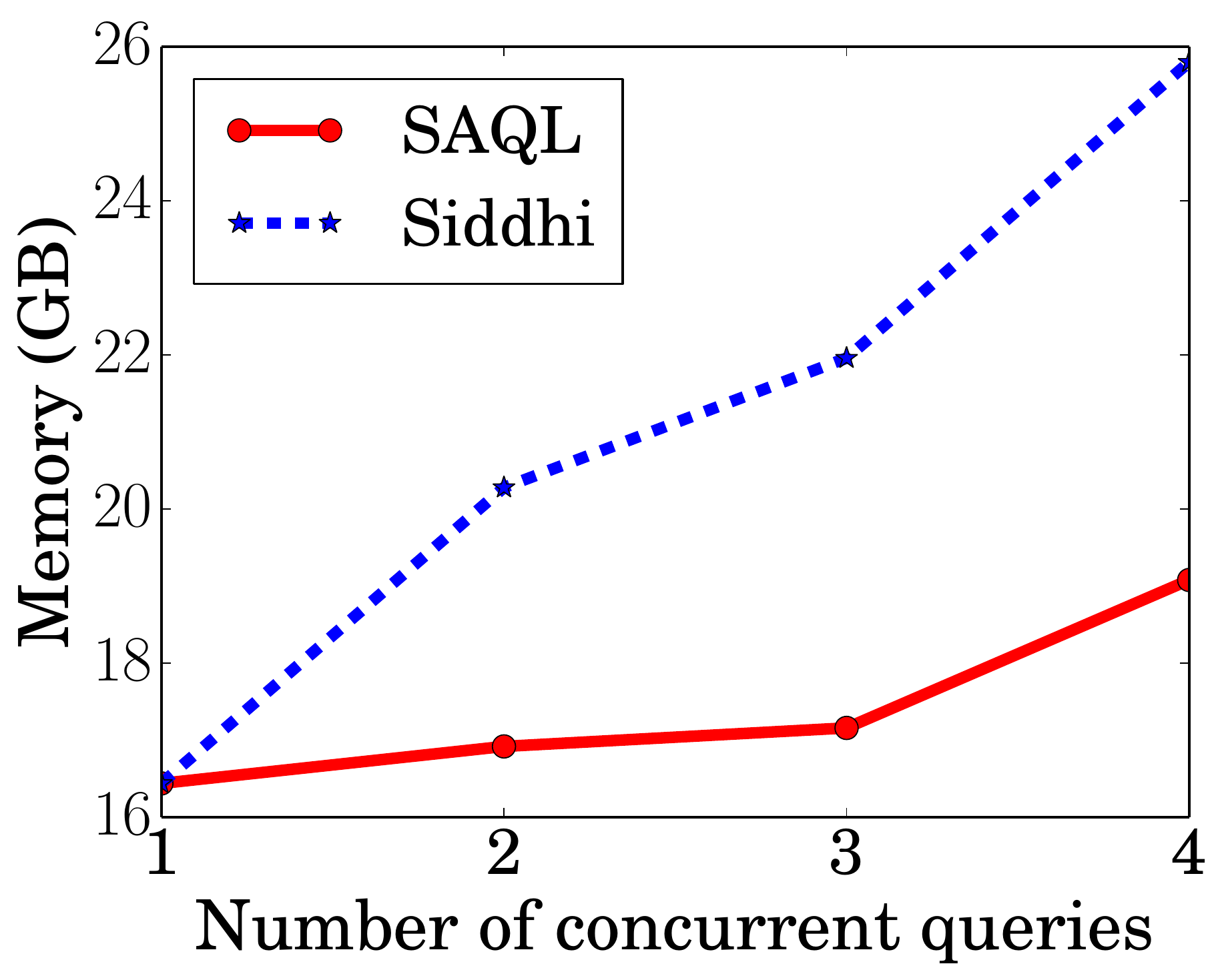}
  \caption{\attNetwork}
\end{subfigure}%
\hfill
\begin{subfigure}[H]{0.25\textwidth}
  \includegraphics[width=\linewidth]{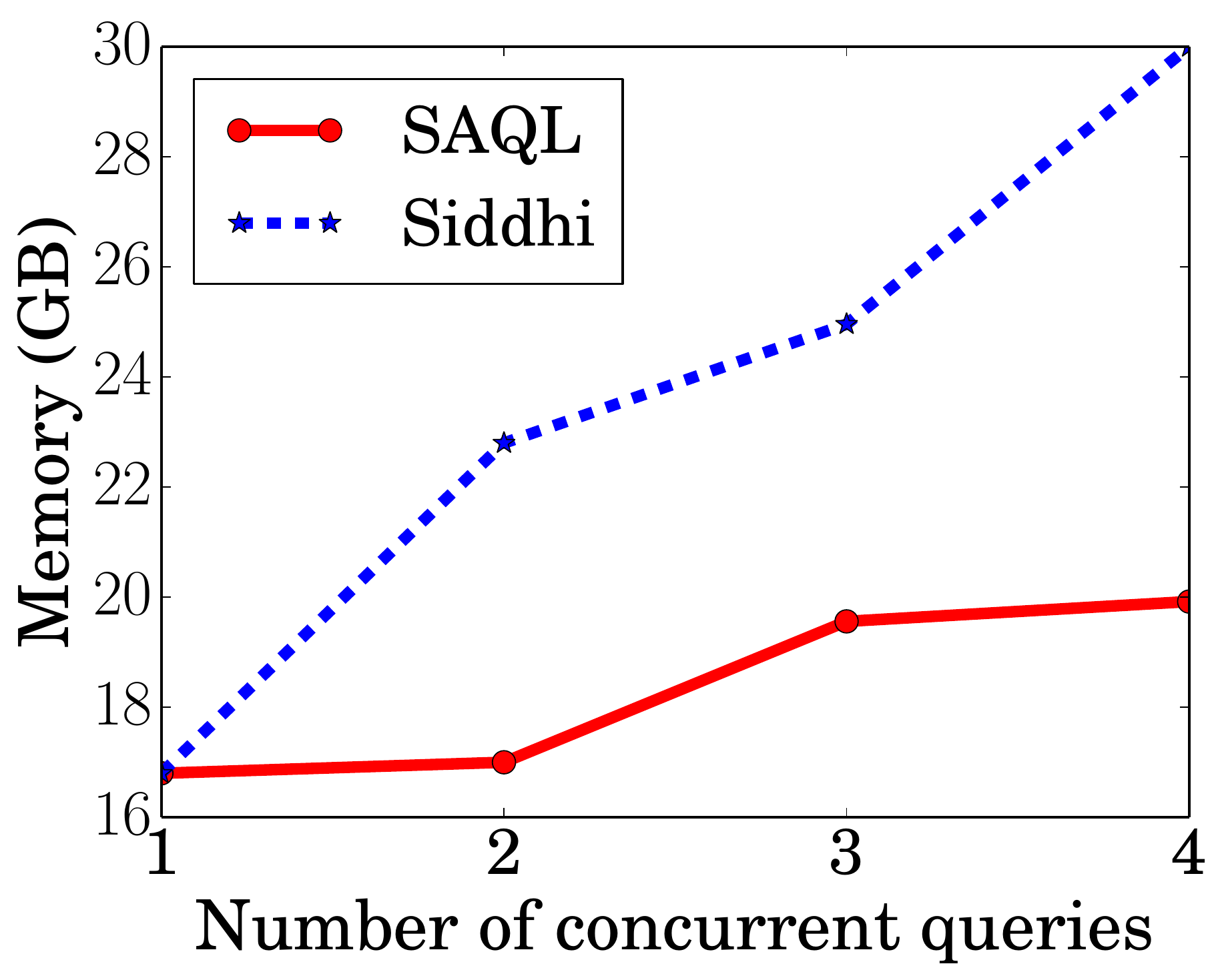}
  \caption{\attProcess}
\end{subfigure}
\caption{\evalWindow}
\label{fig:micro-benchmarks:window}
\vspace*{1ex}
\end{figure*}

\begin{figure*}[!h]
\center
\begin{subfigure}[H]{0.25\textwidth}
  \includegraphics[width=\linewidth]{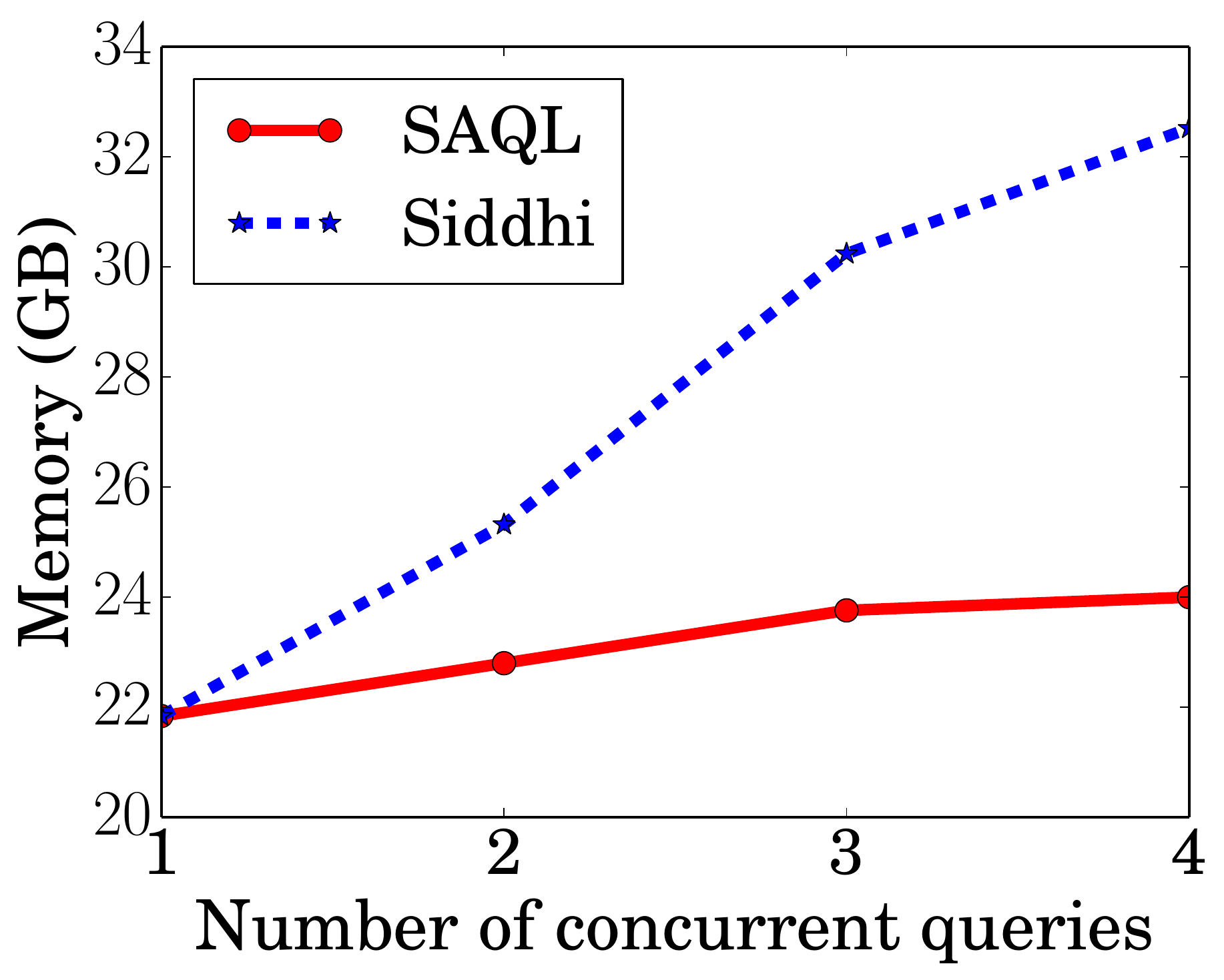}
  \caption{\attFile}
\end{subfigure}%
\hfill
\begin{subfigure}[H]{0.25\textwidth}
  \includegraphics[width=\linewidth]{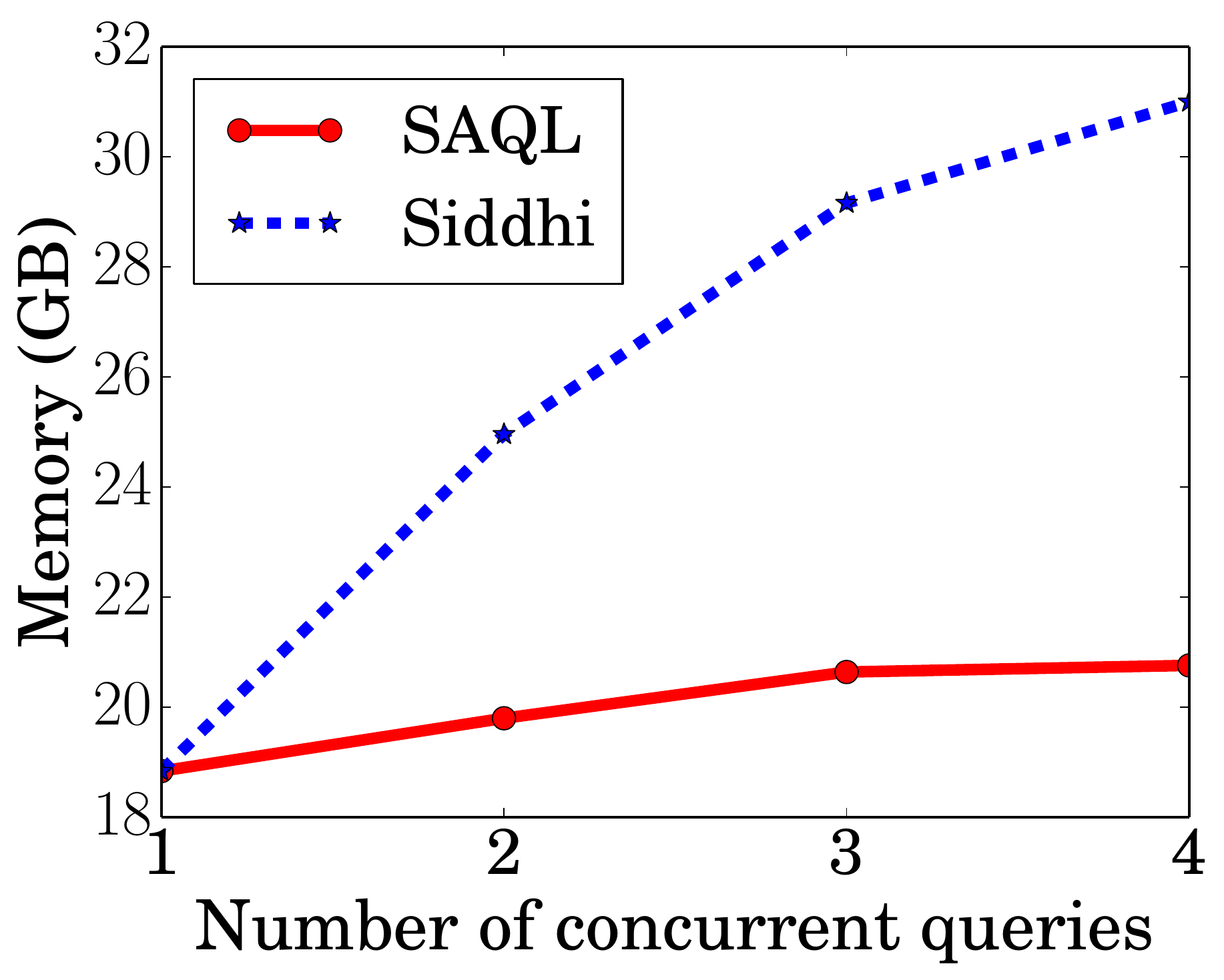}
  \caption{\attBrowser}
\end{subfigure}%
\hfill
\begin{subfigure}[H]{0.25\textwidth}
  \includegraphics[width=\linewidth]{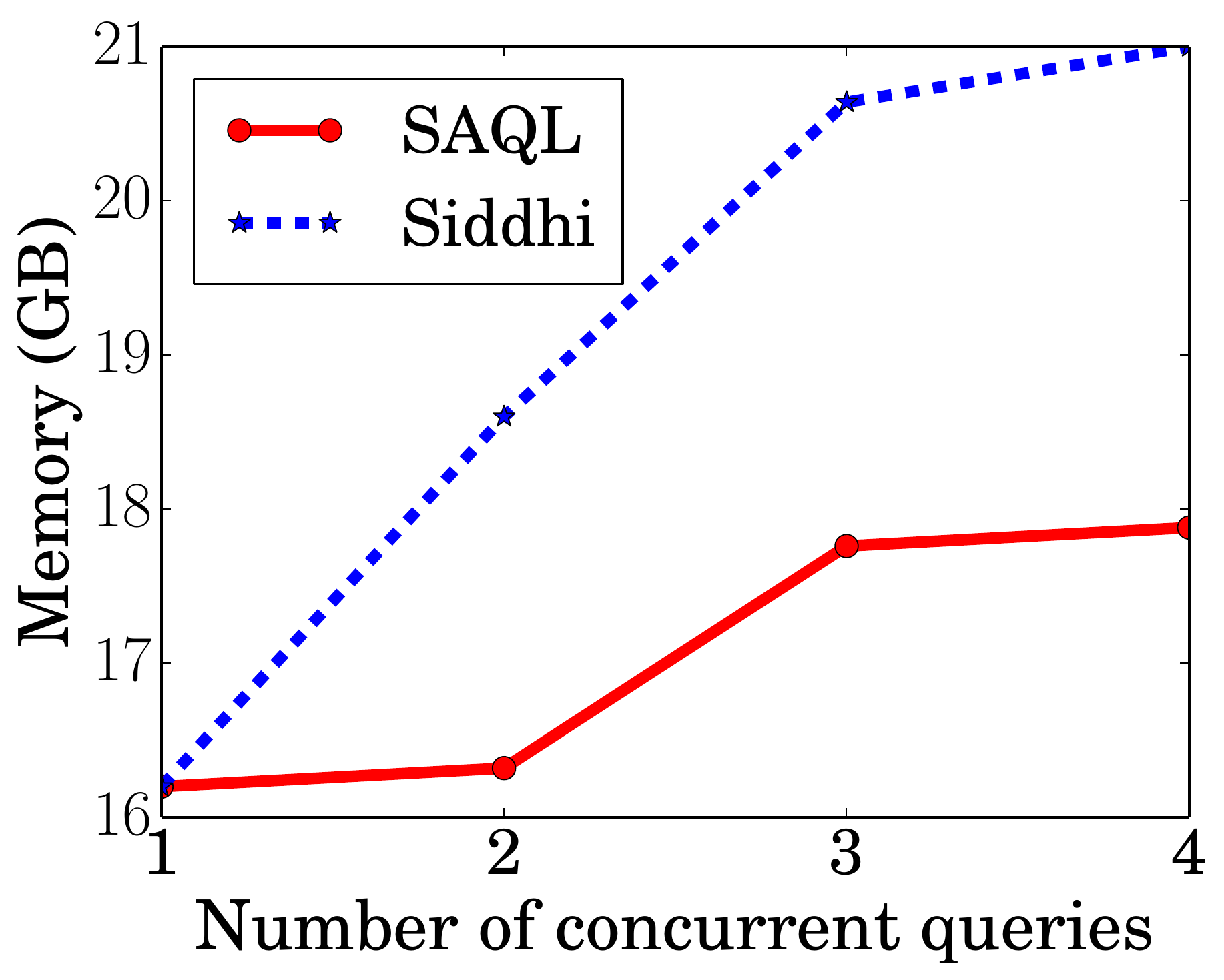}
  \caption{\attNetwork}
\end{subfigure}%
\hfill
\begin{subfigure}[H]{0.25\textwidth}
  \includegraphics[width=\linewidth]{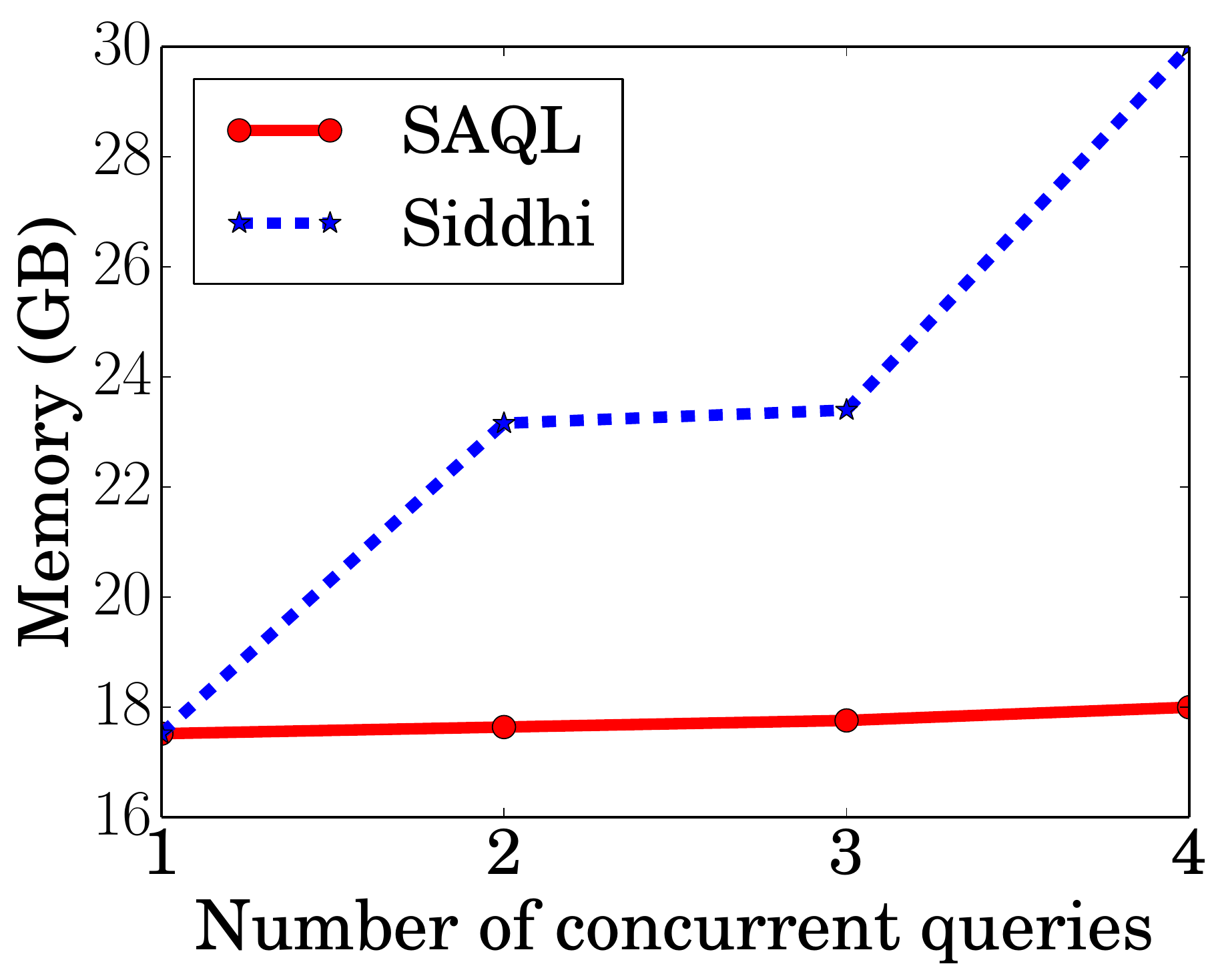}
  \caption{\attProcess}
\end{subfigure}
\caption{\evalAgent}
\label{fig:micro-benchmarks:agentid}
\vspace*{1ex}
\end{figure*}

\begin{figure*}[!h]
\center
\begin{subfigure}[H]{0.25\textwidth}
  \includegraphics[width=\linewidth]{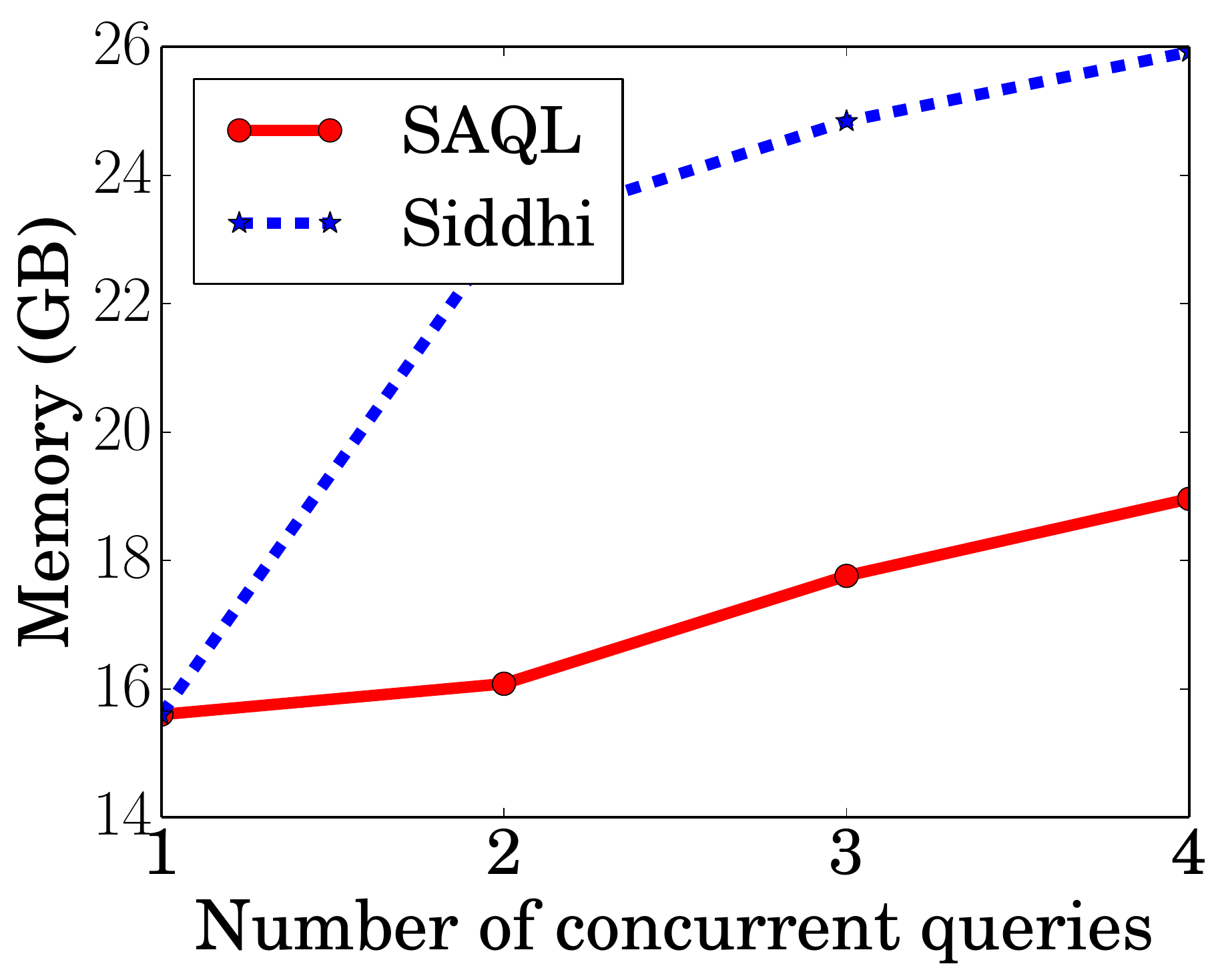}
  \caption{\attFile}
\end{subfigure}%
\hfill
\begin{subfigure}[H]{0.25\textwidth}
  \includegraphics[width=\linewidth]{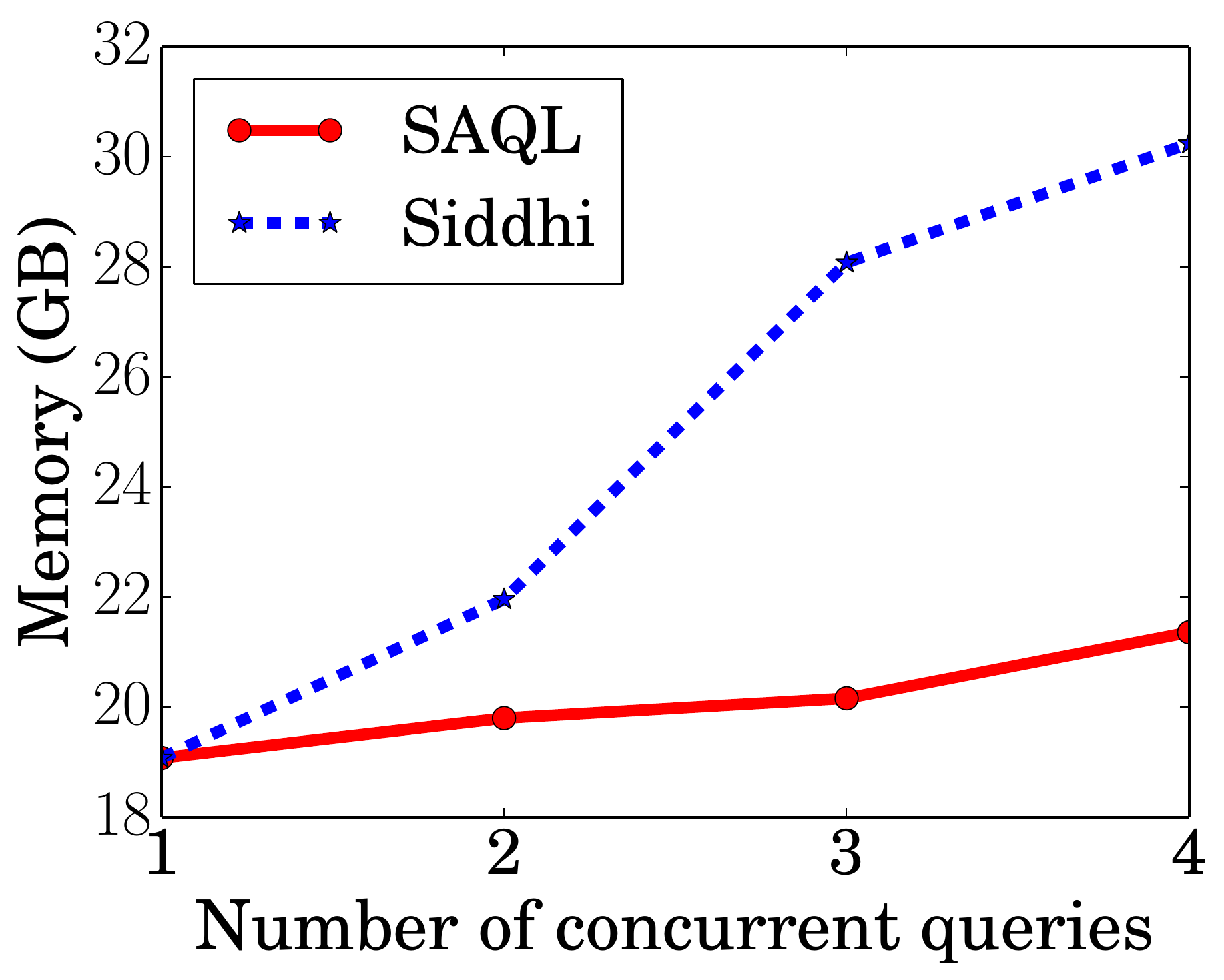}
  \caption{\attBrowser}
\end{subfigure}%
\hfill
\begin{subfigure}[H]{0.25\textwidth}
  \includegraphics[width=\linewidth]{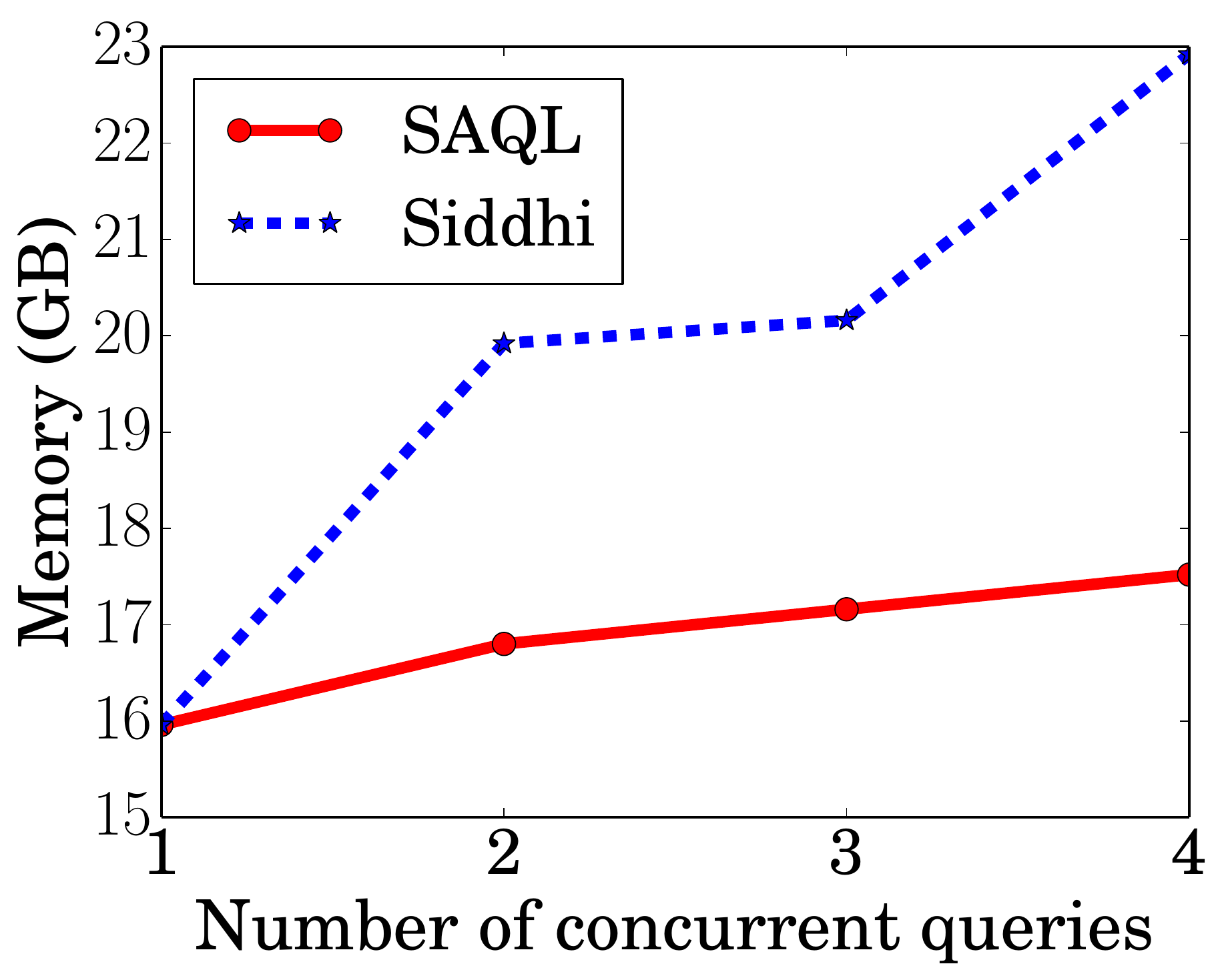}
  \caption{\attNetwork}
\end{subfigure}%
\hfill
\begin{subfigure}[H]{0.25\textwidth}
  \includegraphics[width=\linewidth]{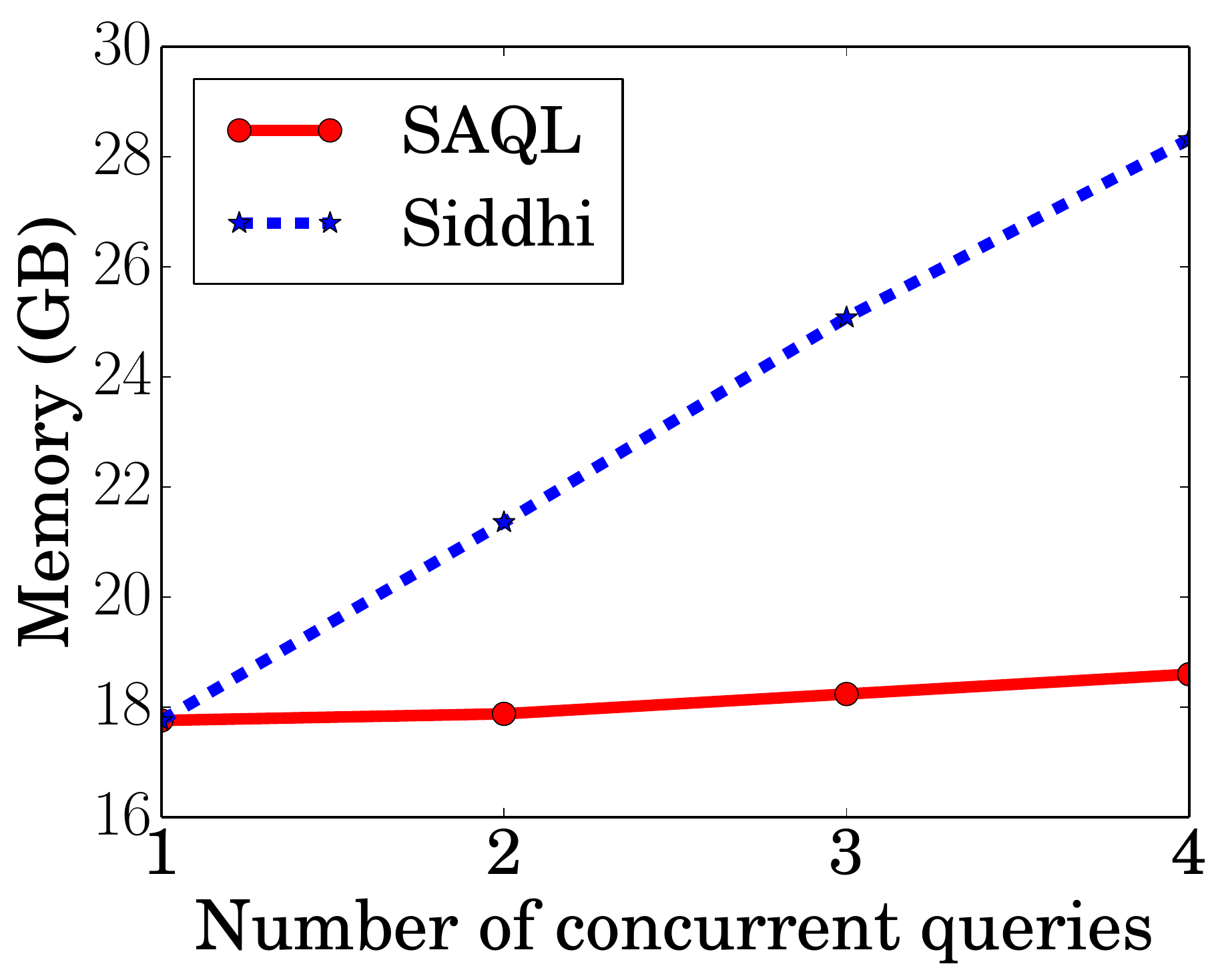}
  \caption{\attProcess}
\end{subfigure}
\caption{\evalState}
\label{fig:micro-benchmarks:state}
\vspace*{1ex}
\end{figure*}

\subsection{Performance Evaluation of Concurrent Query Execution}
\label{subsec:eval-perf}

To evaluate the effectiveness of our query engine (\ie master-dependent-query scheme)
in handling concurrent queries, we construct a micro-benchmark that consists of 64 queries and measure the memory usage during the 
execution.
We select Siddhi~\cite{siddhi}, one of the most popular stream processing and complex event processing engines, for baseline comparison.

\para{Micro-Benchmark Construction}
We construct our micro-benchmark queries by extracting critical attributes from the attacks in \cref{subsubsec:attacks}. 
In particular, we specify the following four \emph{attack categories}:
\begin{itemize}[noitemsep, topsep=1pt, partopsep=1pt, listparindent=\parindent, leftmargin=*]
\item \emph{Sensitive file accesses}: finding processes that access the files \incode{/etc/passwd}, \incode{.ssh/id_rsa}, \incode{.bash_history}, and \incode{/var/log/wtmp}.
\item \emph{Browsers access files}: finding files accessed by the processes \incode{chrome}, \incode{firefox}, \incode{iexplore}, and \incode{microsoftedge}.
\item \emph{Processes access networks}: finding network accesses of the processes \incode{dropbox}, \incode{sqlservr}, \incode{apache}, and \incode{outlook}.
\item \emph{Processes spawn}: finding processes spawn by the processes \incode{/bin/bash}, \incode{/usr/bin/ssh}, \incode{cmd.exe}, and \incode{java}.
\end{itemize}

We also specify the following four \emph{evaluation categories} for query variations, which correspond to the four \emph{optimization dimensions} in \cref{subsec:optimization}: 
\begin{itemize}[noitemsep, topsep=1pt, partopsep=1pt, listparindent=\parindent, leftmargin=*]
\item \emph{Event attributes}: we vary from 1 attribute to 4 attributes. The attributes are chosen from one of the attack categories. The default is 4 attributes.
\item \emph{Sliding window}: we vary from 1 minute to 4 minutes. The default is 1 minute.
\item \emph{Agent ID}: we vary from 1 agent to 4 agents. The default is to avoid the agent ID specification (\ie the query matches all agents).
\item \emph{State aggregation}: we vary from 1 aggregation type to 4 aggregation types, which are chosen from the pool \{\incode{count}, \incode{sum}, \incode{avg}, \incode{max}\}. The default is to avoid the state specification (\ie no states defined).
\end{itemize}

We construct 4 queries for each evaluation category and each attack category. In total, we construct 64 queries for the micro-benchmark. 
For each \dsl query, we construct an equivalent Siddhi query.
Note that unlike \dsl which provides explicit constructs for stateful computation, Siddhi as well as other stream-based query systems~\cite{siddhi,esper,cql,flink}, do not provide the native support for these concepts, making these tools unable to specify advanced anomaly models (\ie time-series anomalies, invariant-based anomalies, outlier-based anomalies). Thus, for the ``state evaluation category'', we only construct Siddhi queries that monitor the same event pattern without stateful computation.
\cref{query:benchmark} shows an example micro-benchmark query for the joint category ``sensitive file accesses \& state aggregation''. 

\begin{lstlisting}[captionpos=b, caption={Example micro-benchmark query}, label={query:benchmark}]
proc p read || write file f["/etc/passwd" || "%.ssh/id_rsa" || "%.bash_history" || "/var/log/wtmp"] as evt #time(1 min)
state ss {
	e1 := count(evt.id)
	e2 := sum(evt.amount)
	e3 := avg(evt.amount)
	e4 := max(evt.amount)
} group by p
return p, ss.e1, ss.e2, ss.e3, ss.e4
\end{lstlisting}

\para{Evaluation Results}
For each evaluation category and each attack category, we vary the number of concurrent queries from 1 to 4 and measure the corresponding memory usage. \cref{fig:micro-benchmarks:attribute,fig:micro-benchmarks:window,fig:micro-benchmarks:agentid,fig:micro-benchmarks:state} show the results.
We observe that:
(1) as the number of concurrent queries increases, the memory usage increases of Siddhi are much higher than the memory usage increases of \dsl in all evaluation settings;
(2) when there are multiple concurrent queries in execution, \dsl require a smaller memory usage than Siddhi in all evaluation settings (30\% average saving when there are 4 concurrent queries).
Such results indicate that the master-dependent-query scheme employed in our query engine is able to save memory usage by sharing the intermediate execution results among dependent queries.
On the contrary, the Siddhi query engine performs data copies, resulting in significantly more memory usage than our query engine.
Note that for evaluation fairness, we use the replayer (\cref{subsec:setup}) to replay a large volume of data in a short period of time. 
Thus, the memory measured in \cref{fig:micro-benchmarks:attribute,fig:micro-benchmarks:window,fig:micro-benchmarks:agentid,fig:micro-benchmarks:attribute} is larger than the memory measured in the case study (\cref{tab:case}), where we use the real-time data streams. Nevertheless, this does not affect the relative improvement of \dsl over Siddhi in terms of memory utilization.

%

\section{Discussion}

\para{Scalability}
The collection of system monitoring data and the execution of \dsl queries can be potentially parallelized with distributed computing.
Parallelizing the data collection involves allocating computing resources (\ie computational nodes) to disjoint sets of enterprise hosts to form sub-streams. 
Parallelizing the \dsl query execution can be achieved through a query-based manner (\ie allocating one computing resource for executing a set of queries over the entire stream), a substream-based manner (\ie allocating one computing resource for executing all compatible queries over a set of sub-streams), or a mixed manner.
Nonetheless, the increasing scale of the deployed environment, the increasing number of submitted queries, and the diversity and semantic dependencies among these queries bring significant challenges to parallel processing. 
Thus, the adaptation of our master-dependent-query scheme to such complicated scenarios is an interesting research direction that requires non-trivial efforts. 
In this work, however, we do not enable distributed computation in our query execution. 
Instead, we collect system monitoring data from multiple hosts, model the data as a single holistic event stream, and execute the queries over the stream in a centralized manner.
Nevertheless, we build our system on top of Siddhi, which can be easily adapted to a distributed mode by leveraging Apache Storm~\cite{siddhidist}. 
Again, we would like to point out that the major focus of our work is to provide a useful interface for investigators to query a broad set of abnormal behaviors from system audit logs, which is orthogonal to the computing paradigms of the underlying stream processing systems.

%

\para{System Entities and Data Reduction}
Our current data model focuses on files, processes, and network connections.
In future work, we plan to expand the monitoring scope by including inter-process communications such as pipes in Linux.
We also plan to incorporate finer granularity system monitoring, such as execution partition to record more precise activities of processes~\cite{mpi,protracer} and in-memory data manipulations~\cite{panda,r2}. 
Such additional monitoring data certainly adds a lot more pressure to the \dsl system, 
and thus more research on data reduction, besides the existing works~\cite{loggc,reduction}, should be explored.

\para{Master-Dependent Query}
Our optimization focuses on the queries that share the pattern matching results and stateful computation results. 
More aggressive sharing could include alerts and even results reported by the alerts,
which we leave for future work.

\para{Anomaly Models}
We admit that while \dsl supports major anomaly models used in commonly observed attacks, there are many more anomaly models that are valuable for specialized attacks.
Our \dsl now allows easy plugins for different clustering algorithms, and we plan to make the system extensible to support more anomaly models by providing interfaces to interact with the anomaly models written in other languages.

\para{Alert Fusion}
Recent security research~\cite{fusion1,fusion2,fusion3} shows promising results in improving detect accuracy using alert fusion that considers multiple alerts.
While this is beyond the scope of this work, our \dsl can be extended with the syntax that supports the specifications of the temporal relationships among alerts.
More sophisticated relationships
would require further design on turning each \dsl query into a module and chaining the modules using various computations.

\section{Related Work}
\label{sec:related}

\para{Audit Logging and Forensics}
Significant progress has been made to leverage system-level provenance for forensic analysis, 
with the focus on generating provenance graphs for attack causality analysis~\cite{mpi,protracer,backtracking,backtracking2,trustkernel,loggc,reduction}. 
Recent work also investigates how to filter irrelevant activities in provenance graphs~\cite{priortracker}
and how to reduce the storage overheads of provenance graphs generated in distributed systems such as data centers~\cite{winnower}.
These systems consider historical logs and their contributions are orthogonal to the contribution of \dsl, which provides a useful and novel interface for investigators to query abnormal behaviors from the stream of system logs. 
Nevertheless, \dsl can be interoperated with these systems to perform causality analysis on the detected anomalies over the concise provenance graphs.

Gao et al.~\cite{aiql} proposed AIQL which enables efficient attack investigation by querying historical system audit logs stored in databases. 
AIQL can be used to investigate the real-time anomalies detected by our \dsl system over the stream of system monitoring data.
Together, these two systems can provide a better defense against advanced cyber attacks.

\para{Security-Related Languages}
There exist domain-specific languages in a variety of security fields that have a well-established corpus of low level
algorithms, such as cryptographic systems~\cite{crypto1,crypto2,oblivm}, secure overlay networks~\cite{mace,networklang}, and network intrusions~\cite{chimera,lambda,HILTI,VAST} and obfuscations~\cite{Marionette:2015}.
These languages are explicitly designed to solve domain specific problems, 
providing specialized constructs for their particular problem domain and eschewing irrelevant features.
In contrast to these languages, the novelty of \dsl focuses on how to specify anomaly models as queries
and how to execute the queries over system monitoring data.
%


\para{Security Anomaly Detection}
Anomaly detection techniques have been widely used in detecting malware~\cite{Hofmeyr1998,staticanalyzer,malwaresystemcall,accessminer}, preventing network intrusion~\cite{trafficaggregation,networkmalware,Portnoy01intrusiondetection},  internal threat detection~\cite{insider}, and attack
prediction~\cite{ai2}.
Rule-based detection techniques characterize normal behaviors of programs through analysis and detect unknown behaviors that have not been observed during the characterization~\cite{502675,Hofmeyr1998}.
Outlier-based detection techniques~\cite{trafficaggregation,networkmalware,Portnoy01intrusiondetection} detect unusual system behaviors based on clustering or other machine learning models.
Unlike these techniques, which focus on finding effective features and building specific models under different scenarios, \dsl provides a unified interface to express anomalies based on domain knowledge of experts.

\para{Complex Event Processing Platforms \& Data Stream Management Systems}
\ac{CEP} platforms, such as Esper~\cite{esper}, Siddhi~\cite{siddhi}, Apache Flink~\cite{flink}, and Aurora~\cite{Abadi:2003:ANM:950481.950485} match continuously incoming events against a pattern. 
Unlike traditional database management systems where a query is executed on the stored data,
CEP queries are applied on a potentially infinite stream of data,
and all data that is not relevant to the query is immediately discarded.
These platforms provide their own domain-specific languages that can compose patterns of complex events with the support of sliding windows.  Wukong+S~\cite{streamanalytics4} builds a stream querying platform that can query both the stream data and stored data. 
Data stream management systems~\cite{dsms}, such as CQL~\cite{cql}, manage multiple data streams and provide a query language to process the data over the stream.
These \ac{CEP} platforms are useful in managing large streams of data. Thus, they can be used as a management infrastructure for our approach. 
However, these \ac{CEP} systems alone do not provide language constructs to support stateful computation in sliding windows,
and thus lack the capability to express stateful anomaly models as our system does.

\para{Stream Computation Systems}
Stream computation systems allow users to compute various metrics based on the stream data. These systems include Microsoft StreamInsight~\cite{StreamInsight}, MillWheel~\cite{MillWheel}, Naiad~\cite{Naiad}, and Puma~\cite{sigmod16realtime}. These systems normally provide a good support for stateless computation (e.g., data aggregation). 
However, they do not support stateful anomaly models as our \dsl system does, which are far more complex than data aggregation.

\para{Other System Analysis Languages}
Splunk~\cite{splunk} and  Elasticsearch~\cite{elasticsearch} are platforms that automatically parse general application logs,
and provide a keyword-based search language to filter entries of logs.
OSQuery~\cite{osquery,osquerysec} allows analysts to use SQL queries to probe the real-time system status.
However, these systems and the languages themselves cannot support anomaly detection and do not support stateful computation in sliding windows.
Other languages, such as Weir~\cite{weir} and StreamIt~\cite{streamit}, focus on monitoring the system performance, 
and lack support for expressing anomaly models.


\section{Conclusion}
\label{sec-conclusions}

We have presented a novel stream-based query system that takes a real-time event feed aggregated from different hosts under monitoring,
and provides an anomaly query engine that  checks the event stream against the queries submitted by security analysts to detect anomalies in real-time.
Our system provides a \emph{domain-specific language}, \dsl, which is specially designed to facilitate the task of expressing anomalies based on domain knowledge.
\dsl provides the constructs of event patterns to easily specify relevant system activities and their relationships,
and the constructs to perform stateful computation by defining states in sliding windows and accessing historical states to compute anomaly models.
With these constructs, \dsl allows security analysts to express models for (1) \emph{rule-based anomalies}, (2) \emph{time-series anomalies},
(3) \emph{invariant-based anomalies},
and (4) \emph{outlier-based anomalies}.
Our evaluation results on 17 attack queries and 64 micro-benchmark queries show that the \dsl system has a low alert detection latency and a high system throughput, and is more efficient in memory utilization than the existing stream processing systems.

%

\section{Acknowledgements}
We would like to thank the anonymous reviewers and our shepherd, Prof. Adam Bates, for their insightful feedback in finalizing this paper.
This work was partially supported by the National Science Foundation under grants CNS-1553437 and CNS-1409415.
Any opinions, findings, and conclusions made in this material are those of the authors and do not necessarily reflect the views of the funding agencies.


{\footnotesize \bibliographystyle{acm}
\bibliography{ref}}



\vspace{2ex}

\appendix

\noindent\textbf{\large Appendix}

\section{\dsl Queries in Attack Cases Study}

We present the 17 \dsl queries that we construct in the case study, which are used detect the four major types of attack behaviors (\cref{subsubsec:attacks}). For privacy purposes, we anonymize the IP addresses and the agent IDs in the presented queries.

\subsection{APT Attack}
\label{app:apt}

\begin{lstlisting}[captionpos=b, caption={apt-c1}, label={query:c1}]
proc p1["%smtp%"] read||write ip i1[srcip="XXX" && srcport=25 && protocol=6] as evt1[agentid = XXX] // mail server, SMTP connection from the router to the mail server
proc p2["%imap%"] read||write ip i2[srcip="XXX" && srcport=143 && dstip="XXX" && dstport=51962 && protocol=6] as evt2[agentid = XXX] // mail server, IMAP connection from the mail server to the client
proc p3["%outlook%"] read||write ip i3[srcip="XXX" && srcport=51960 && dstip="XXX" && dstport=143 && protocol=6] as evt3[agentid = XXX] // windows client, client's outlook reads email data
with evt1 -> evt2 -> evt3
return p1, i1, p2, i2, p3, i3, evt1.starttime, evt2.starttime, evt3.starttime
\end{lstlisting}

\begin{lstlisting}[captionpos=b, caption={apt-c2}, label={query:c2}]
agentid = XXX // windows client
proc p1["%outlook.exe"] start proc p2["%excel.exe"] as evt1 // outlook starts excel
proc p2 start proc p3["%java.exe"] as evt2 // excel starts malware (java) process
proc p3 start proc p4["%notepad.exe"] as evt3 // malware (java) starts notepad
proc p4 read||write ip i1["XXX"] as evt4 // notepad connects to the attacker host
with evt1 -> evt2 -> evt3 -> evt4
return p1, p2, p3, p4, i1, evt1.starttime, evt2.starttime, evt3.starttime, evt4.starttime
\end{lstlisting}

\begin{lstlisting}[captionpos=b, caption={apt-c3}, label={query:c3}]
agentid = XXX // windows domain controller
proc p1 read || write ip i1[srcport=445 && dstip="XXX"] as evt1 // attacker penetrates to the DC host using psexec protocol
proc p2["%powershell.exe"] write file f1["%gsecdump%"] as evt2 // attacker transfers the DB cracking tool gsecdump.exe
proc p3["%cmd.exe"] start proc p4["%gsecdump%"] as evt3 // attacker executes gsecdump.exe to dump DB administrator credentials
with evt1 -> evt2 -> evt3
return p1, i1, p2, f1, p3, p4, evt1.starttime, evt2.starttime, evt3.starttime
\end{lstlisting}

\begin{lstlisting}[captionpos=b, caption={apt-c4}, label={query:c4}]
agentid = XXX // db server
proc p1["%sqlservr.exe"] read||write ip i1[srcip="XXX" && srcport=1433 && dstip="XXX" && dstport=52038 && protocol=6] as evt1 // attacker connects to the SQL server using DB administrator credentials
proc p1 start proc p2["%cmd.exe"] as evt2 // SQL server starts cmd
proc p2 read || write file f1["%hwvun.vbs"] as evt3 // cmd writes malware sbblv.exe
proc p3["%cscript.exe"] write file f2["%sbblv.exe"] as evt4
proc p4["%sbblv.exe"] start ip i2[srcip="XXX" && srcport=61060 && dstip="XXX" && dstport=443 && protocol=6] as evt5 // malware connects back to the attacker host
with evt1 -> evt2 -> evt3 -> evt4 -> evt5
return p1, i1, p2, f1, p3, f2, p4, i2, evt1.starttime, evt2.starttime, evt3.starttime, evt4.starttime, evt5.starttime
\end{lstlisting}

\begin{lstlisting}[captionpos=b, caption={apt-c5}, label={query:c5}]
agentid = XXX // db server
proc p1["%cmd.exe"] start proc p2["%osql.exe"] as evt1 // attacker executes osql.exe on the sql server
proc p3["%sqlservr.exe"] write file f1["%backup1.dmp"] as evt2 // attacker dumps the DB content
proc p4["%sbblv.exe"] read file f1 as evt3 // malware reads the dump
proc p4 read || write ip i1[dstip="XXX"] as evt4 // malware transfers the dump to the attacker
with evt1 -> evt2 -> evt3 -> evt4
return p1, p2, p3, f1, p4, i1, evt1.starttime, evt2.starttime, evt3.starttime, evt4.starttime, evt4.amount
\end{lstlisting}

\begin{lstlisting}[captionpos=b, caption={apt-c2-invariant}, label={query:c2:invariant}]
proc p1["%excel.exe"] start proc p2 as evt #time(5 second)
state ss {
	set_proc := set(p2.exe_name)
} group by p1, evt.agentid
invariant[100][offline] {
	a := empty_set
	a = a union ss.set_proc
}
alert |ss.set_proc diff a| > 0
return p1, evt.agentid, ss.set_proc
\end{lstlisting}

\begin{lstlisting}[captionpos=b, caption={apt-c5-timeseries}, label={query:c5:timeseries}]
agentid = XXX // db server
proc p write ip i as evt #time(10 min)
state[3] ss {
	avg_amount := avg(evt.amount)
} group by p
alert (ss[0].avg_amount > (ss[0].avg_amount + ss[1].avg_amount + ss[2].avg_amount) / 3) && (ss[0].avg_amount > 10000)
return p, ss[0].avg_amount, ss[1].avg_amount, ss[2].avg_amount
\end{lstlisting}

\begin{lstlisting}[captionpos=b, caption={apt-c5-outlier}, label={query:c5:outlier}]
agentid = XXX// db server
proc p write ip i as evt #time(1 min)
state ss {
	avg_amount := avg(evt.amount)
} group by p
cluster(points=all(ss.avg_amount), distance="ed", method="DBSCAN(1000, 5)")
alert cluster.outlier && ss.avg_amount > 1000000
return p, ss.avg_amount
\end{lstlisting}

\subsection{SQL Injection Attack}
\label{app:sql}

\begin{lstlisting}[captionpos=b, caption={sql-injection}, label={query:sql}]
agentid = XXX // sqlserver host
proc p["%sqlservr.exe"] read || write ip i as evt #time(10 min)
state ss {
	amt := sum(evt.amount)
} group by i.dstip
cluster(points=all(ss.amt), distance="ed", method="DBSCAN(100000, 5)")
alert cluster.outlier && ss.amt > 1000000
return i.dstip, ss.amt
\end{lstlisting}

\subsection{Bash Shellshock Command Injection Attack}
\label{app:shellshock}

\begin{lstlisting}[captionpos=b, caption={shellshock}, label={query:shellshock}]
proc p1["%apache2%"] start proc p2 as evt #time(10 s)
state ss {
	set_proc := set(p2.exe_name)
} group by p1
invariant[10][offline] {
	a := empty_set // invariant init
	a = a union ss.set_proc //invariant update
}
alert |ss.set_proc diff a| > 0
return p1, ss.set_proc
\end{lstlisting}

\subsection{Suspicious System Behaviors}
\label{app:suspicious}

\begin{lstlisting}[captionpos=b, caption={dropbox}, label={query:dropbox}]
proc p["%dropbox%"] start ip i as evt
return p, i, evt.agentid, evt.starttime, evt.endtime
\end{lstlisting}

\begin{lstlisting}[captionpos=b, caption={command-history}, label={query:history}]
proc p read || write file f["%.viminfo" || "%.bash_history" || "%.zsh_history" || "%.lesshst" || "%.pgadmin_histoqueries" || "%.mysql_history"] as evt
return p, f, evt.agentid, evt.starttime, evt.endtime
\end{lstlisting}

\begin{lstlisting}[captionpos=b, caption={password}, label={query:password}]
proc p read || write file f["/etc/passwd"] as evt
return p, f, evt.agentid, evt.starttime, evt.endtime
\end{lstlisting}

\begin{lstlisting}[captionpos=b, caption={login-log}, label={query:login}]
proc p write file f["/var/log/wtmp" || "/var/log/lastlog"] as evt
return p, f, evt.agentid, evt.starttime, evt.endtime
\end{lstlisting}

\begin{lstlisting}[captionpos=b, caption={sshkey}, label={query:sshkey}]
proc p read || write file f["%.ssh/id_rsa" || "%.ssh/id_dsa"] as evt
return p, f, evt.agentid, evt.starttime, evt.endtime
\end{lstlisting}

\begin{lstlisting}[captionpos=b, caption={usb}, label={query:usb}]
proc p read || write file f[bustype = "USB"] as evt
return p, f, evt.agentid
\end{lstlisting}

\begin{lstlisting}[captionpos=b, caption={ipfreq}, label={query:ipfreq}]
proc p start ip ipp #time(1 min)
group by p
alert freq > 100
return p, count(ipp) as freq
\end{lstlisting}

\end{document}